\documentclass[reprint,twocolumn,notitlepage,preprintnumbers,nofootinbib,amsmath,amssymb,aps,prd,floatfix,superscriptaddress,longbibliography,svgnames]{revtex4-1}

\usepackage{amsmath}
\usepackage{amsfonts}
\usepackage{amssymb}
\usepackage{graphicx}
\usepackage{bm}
\usepackage{subfigure}
\usepackage{url}
\usepackage[hyperindex]{hyperref}
\usepackage{color}
\usepackage[ddmmyy,24hr]{datetime}
\usepackage{bigdelim}
\usepackage{booktabs}
\usepackage{dcolumn}
\usepackage{multirow}
\usepackage{subfigure}
\usepackage{physics}
\usepackage{cancel}
\usepackage{stackrel}
\usepackage{paralist}
\usepackage{xspace}
\usepackage{slashed}
\usepackage{cancel}
\usepackage{todonotes}
\usepackage{enumerate}
\usepackage{float}
\usepackage{fullpage}
\usepackage{ulem}
\usepackage{wasysym}
\usepackage{comment}
\usepackage{bbold}
\usepackage{hyperref}
\newcommand{\cenns}{CE$\nu$NS }
\usepackage{xcolor}
\usepackage{orcidlink}

\usepackage{stackengine}

\usepackage{feynmp-auto}
\newcommand{\nua}[1]{\ensuremath{\rlap{\kern-2.5pt\ensuremath{\overset{\scriptscriptstyle(-)}{\phantom{\nu}}}}{\ensuremath{{\nu}_{#1}}}}}

\usepackage{enumitem}

\newcommand{\cevns}{CE$\nu$NS }

\begin{document}

\title{Joint analysis of reactor and accelerator CE$\nu$NS data on germanium: implications for the Standard Model and nuclear physics}

\author{M. Atzori Corona \orcidlink{0000-0001-5092-3602}}
\email{mcorona@roma2.infn.it}
\affiliation{Istituto Nazionale di Fisica Nucleare (INFN), Sezione di Roma Tor Vergata, Via della Ricerca Scientifica, I-00133 Rome, Italy}

\author{M. Cadeddu \orcidlink{0000-0002-3974-1995}}
\email{matteo.cadeddu@ca.infn.it}
\affiliation{Istituto Nazionale di Fisica Nucleare (INFN), Sezione di Cagliari,
	Complesso Universitario di Monserrato - S.P. per Sestu Km 0.700,
	09042 Monserrato (Cagliari), Italy}

\author{N. Cargioli \orcidlink{0000-0002-6515-5850}}
\email{nicola.cargioli@ca.infn.it}
\affiliation{Istituto Nazionale di Fisica Nucleare (INFN), Sezione di Cagliari,
	Complesso Universitario di Monserrato - S.P. per Sestu Km 0.700,
	09042 Monserrato (Cagliari), Italy}

    \author{G. Co' \orcidlink{0000-0002-9613-5211}}
\email{giampaolo.co@le.infn.it}
\affiliation{Dipartimento di Matematica e Fisica ``E. De Giorgi''
Universit\`a del Salento.\\
Istituto Nazionale di Fisica Nucleare (INFN), Sezione di Lecce\\
via Arnesano, 73100 Lecce, Italy
}

\author{F. Dordei \orcidlink{0000-0002-2571-5067}}
\email{francesca.dordei@cern.ch}
\affiliation{Istituto Nazionale di Fisica Nucleare (INFN), Sezione di Cagliari,
	Complesso Universitario di Monserrato - S.P. per Sestu Km 0.700,
	09042 Monserrato (Cagliari), Italy}

\author{C. Giunti \orcidlink{0000-0003-2281-4788}}
\email{carlo.giunti@to.infn.it}
\affiliation{Istituto Nazionale di Fisica Nucleare (INFN), Sezione di Torino, Via P. Giuria 1, I--10125 Torino, Italy}

\date{\dayofweekname{\day}{\month}{\year} \ddmmyydate\today, \currenttime}

\begin{abstract}
This work presents the first comprehensive joint analysis of all available Coherent Elastic Neutrino-Nucleus Scattering
(CE$\nu$NS) data on germanium: 
those observed at the Spallation Neutron Source (SNS) by the COHERENT collaboration 
and those of the nuclear reactors revealed by the CONUS+ experiment using germanium detectors.
In addition to COHERENT and CONUS+, we incorporate reactor data from TEXONO and $\nu$GeN, thereby enhancing both the statistical significance and the systematic reliability of our study.
We provide state-of-the-art determinations of key nuclear physics and Standard Model parameters, including the neutron root-mean-square (rms) radius of germanium nuclei, the weak mixing angle, and the neutrino charge radius.
The observed tension of about $2 \sigma$
between the COHERENT germanium measurement and the Standard Model prediction motivates a detailed reassessment of the theoretical cross-section. In particular, we examine the impact of nuclear form factors and uncertainties in the nuclear radius, as well as the potential influence of a systematic shift in the neutrino flux normalisation at the SNS.
Our results highlight the reliability of CE$\nu$NS as a precision tool, reinforced by the complementarity of different experimental inputs, and lay the groundwork for future advances in the field.

\end{abstract}

\maketitle  
%\newpage

\section{Introduction}

Coherent elastic neutrino-nucleus scattering (CE$\nu$NS), first predicted in 1974~\cite{Freedman:1973yd}, has emerged in recent years as a powerful 
low-energy probe for both testing the Standard Model (SM) and searching for phenomena beyond it~\cite{Cadeddu:2023tkp,AtzoriCorona:2025xwr}. This process was first observed in 2017 by the COHERENT collaboration~\cite{COHERENT:2017ipa,COHERENT:2018imc, COHERENT:2021xmm} using a CsI detector exposed to the stopped-pion neutrino source at the Spallation Neutron Source (SNS), and subsequently confirmed using a liquid argon target~\cite{COHERENT:2020iec}. These landmark results opened the door to precision neutrino physics in the low-energy regime~\cite{DeRomeri:2024hvc,Akimov:2024lnl,DeRomeri:2024iaw,Majumdar:2024dms,Pandey:2023arh,Coloma:2023ixt,AristizabalSierra:2024nwf, Cadeddu:2017etk, Cadeddu:2018dux, Cadeddu:2019eta, Cadeddu:2020lky, Cadeddu:2018izq, Cadeddu:2020nbr, Cadeddu:2021ijh, AtzoriCorona:2022moj,AtzoriCorona:2022qrf,AtzoriCorona:2023ktl,AtzoriCorona:2024rtv,Coloma:2017ncl,Liao:2017uzy,Lindner:2016wff,Giunti:2019xpr,Denton:2018xmq,AristizabalSierra:2018eqm,Miranda:2020tif,Banerjee:2021laz,Papoulias:2019lfi,Denton:2018xmq,AristizabalSierra:2018eqm,Papoulias:2017qdn,Dutta:2019nbn,Abdullah:2018ykz,Ge:2017mcq,Miranda:2021kre,Flores:2020lji, Farzan:2018gtr, Brdar:2018qqj}, stimulating a new generation of CE$\nu$NS experiments across various platforms.

A major step forward has now been achieved with the first observation of CE$\nu$NS on a germanium target with an active mass of $(10.66\pm0.09)\;\rm{kg}$ at the SNS, as reported by the COHERENT collaboration~\cite{COHERENT:2025vuz}. This marks the first CE$\nu$NS detection from a pion-decay-at-rest neutrino source on a high-purity germanium detector, adding a crucial third nuclear target to the experimental landscape. This result has been implemented with
the recent CE$\nu$NS observations by the CONUS+~\cite{Ackermann:2025obx} collaboration using reactor antineutrinos on a similar germanium target, representing the first unambiguous CE$\nu$NS detection at a reactor site.

These two independent observations on germanium, realised under very different neutrino flux conditions, namely prompt, pulsed neutrinos from pion decay at the SNS versus steady, lower-energy antineutrinos from a commercial reactor, provide a unique opportunity for a simultaneous analysis. 
In reactor-based CE$\nu$NS experiments, such as those conducted by CONUS+ but also the recent intriguing constraints posed by the TEXONO~\cite{TEXONO:2024vfk} and $\nu$GeN~\cite{nuGeN:2025mla} collaborations, the antineutrino energies are limited to a few MeV. At these low energies, the momentum transfer is small enough that the nuclear form factor remains essentially flat and close to unity. As a result, the scattering cross section becomes largely insensitive to nuclear structure effects, offering a cleaner experimental environment for probing fundamental SM parameters, neutrino properties and possible non-standard interactions. 

On the other hand, CE$\nu$NS measurements at the SNS involve neutrinos in the tens-of-MeV range, where the finite momentum transfer begins to probe the nuclear interior, introducing sensitivity to
nuclear models and related uncertainties. 
Moreover, while reactor data is only sensitive to the electronic flavour, accelerator data offers a unique opportunity to probe both electron and muon neutrino flavours.

This complementarity between reactor and accelerator sources is especially valuable when considering data from a common target such as germanium. 
We present 
the first combined analysis of CE$\nu$NS results on germanium, integrating data from both SNS and reactor-based experiments. We provide state-of-the-art determinations of key nuclear physics parameters such as the neutron root-mean-square (rms) radius of germanium nuclei and fundamental SM quantities, including the weak mixing angle, and the neutrino charge radius.

\section{Theory}
\label{sec:theory}

The \cenns differential cross section as a function of nuclear recoil energy \(T_\mathrm{nr}\) for a neutrino \(\nu_\ell\) (\(\ell = e, \mu, \tau\)) scattering off a nucleus \(\mathcal{N}\) with $Z$ protons and $N$ neutrons is~\cite{Cadeddu:2023tkp}\footnote{In this work we use natural units such that $\hbar=c=1$.}
\begin{equation}
    \dfrac{d\sigma_{\nu_{\ell}\text{-}\mathcal{N}}}{d T_\mathrm{nr}} = 
    \dfrac{G_{\text{F}}^2 M}{\pi} 
    \left( 1 - \dfrac{M T_\mathrm{nr}}{2 E^2} \right)
    \left( Q^{V}_{\ell, \mathrm{SM}} \right)^2,
    \label{eq:cexsec}
\end{equation}
where \(G_{\text{F}}\) is the Fermi constant, \(E\) the neutrino energy, \(M\) the nuclear mass\footnote{For the isotopic composition of germanium, we use the values from Ref.~\cite{BerglundWieser+2011+397+410}.} and the weak nuclear charge is
\begin{equation}
    Q^{V}_{\ell, \mathrm{SM}} = \left[ g_{V}^{p}(\nu_\ell) Z F_Z(|\vec{q}|^2) + g_{V}^{n} N F_N(|\vec{q}|^2) \right].
    \label{eq:weakcharge}
\end{equation}
We indicate with $F_Z$ and $F_N$ the proton and neutron form factors of the nucleus and with $g_{V}^{n}$ and $g_{V}^{p}$ the coefficients which 
quantify the weak neutral-current interactions of neutrons and protons. In the SM, the values of these coefficients are~\cite{AtzoriCorona:2024rtv}
\begin{align}
g_{V}^{p}(\nu_{e}) &= 0.0379,\, g_{V}^{p}(\nu_{\mu}) = 0.0297, \\
g_{V}^{p}(\nu_{\tau}) &= 0.0253,\, g_{V}^{n} = -0.5117, 
\end{align}
when taking into account radiative corrections in the $\overline{\mathrm{MS}}$ scheme~\cite{AtzoriCorona:2023ktl,Erler:2013xha,PhysRevD.110.030001}. 
The flavor dependence of $g_{V}^{p}(\nu_\ell)$ is due to the neutrino charge radii (CR), which represent the only non-zero electromagnetic properties of neutrinos in the SM~\cite{Giunti:2024gec}. 
The SM neutrino CR prediction can be written as~\cite{Bernabeu:2000hf,Bernabeu:2002nw}
\begin{equation}
\langle{r}_{\nu_{\ell}}^2\rangle_{\text{SM}}
=
-
\frac{G_{\text{F}}}{2\sqrt{2}\pi^2}
\left[
3 - 2 \ln\left(\frac{m_{\ell}^2}{m_W^2}\right)
\right],
\label{eq:cr-sm}
\end{equation}
where $m_{W}$ is the $W$ boson mass, $m_{\ell}$ is the mass of the charged lepton $\ell = e, \mu, \tau$. 
The SM values of the neutrino CR of interest for this work are:
\begin{align}
\langle{r}_{\nu_{e}}^2\rangle_{\text{SM}} &\simeq -0.83 \times 10^{-32} \, \text{cm}^2, \label{eq:cr-e} \\
\langle{r}_{\nu_{\mu}}^2\rangle_{\text{SM}} &\simeq -0.48 \times 10^{-32} \, \text{cm}^2. \label{eq:cr-mu}
\end{align}

The proton, \(F_Z(|\vec{q}|^2)\), and neutron, \(F_N(|\vec{q}|^2)\), nuclear form factors (FFs) in Eq.~(\ref{eq:weakcharge}) encode the dependence of the process on the nuclear structure, and are defined as the Fourier transform of the corresponding nucleon density
distributions in the nucleus, $\rho_{Z(N)}$, respectively. Their effect becomes more relevant for increasing momentum transfers, \( |\vec{q}|\simeq\sqrt{2MT_{\rm{nr}}} \)~\cite{Cadeddu:2017etk} leading to a suppression of the full coherence~\cite{AtzoriCorona:2023ktl} in the \cenns process. As stated in the introduction, while the FFs are crucial ingredients in the interpretation of COHERENT data, in the low-energy regime of reactor experiments, the FF of both protons and neutrons is practically equal to unity, making the particular choice of the parameterisation almost irrelevant.
We employ the analytical Helm parameterisation to describe both proton and neutron FFs~\cite{Helm:1956zz}. 
This parameterisation is practically equivalent to other two well-known parameterisations, i.e., the symmetrised Fermi~\cite{Piekarewicz:2016vbn} and Klein-Nystrand~\cite{Klein:1999qj} ones.\\
The Helm form factor depends on the rms radius of protons and neutrons. 
While the proton rms radii can be obtained from the charge radius~\cite{Cadeddu:2020lky}, which is extracted from muonic atom spectroscopy and electron scattering data~\cite{Fricke:1995zz,Fricke2004,Angeli:2013epw}, the neutron rms radii lack precise measurements for the target nuclei employed in \cenns experiments. Therefore, some assumptions need to be made.
\begin{table}[h!]
\centering
\renewcommand{\arraystretch}{1.5}
\resizebox{\columnwidth}{!}
{
\begin{tabular}{c|c|c|c||c|c|c}

 & \multicolumn{3}{c||}{$R_p$ [fm]} & \multicolumn{3}{c}{$R_n$ [fm]} \\
\cline{2-7}
 & Ref.~\cite{Fricke:1995zz} & D1S~\cite{Berger:1991zza} & D1M~\cite{Chabanat:1997un} & D1S~\cite{Berger:1991zza} & D1M~\cite{Chabanat:1997un}  & NSM~\cite{Hoferichter:2020osn}\\
\hline
 $^{70}$Ge & 4.055(1) & 3.99  & 3.93 & 4.02 & 3.97 & 4.14\\
\hline
 $^{72}$Ge & 4.073(1) & 4.01  & 3.96 & 4.07 & 4.02 & 4.20 \\
 \hline
 $^{73}$Ge & 4.079(1) & 4.03 & 3.97 & 4.10 & 4.04 & 4.22\\
 \hline
 $^{74}$Ge & 4.091(1) & 4.05 & 3.98 & 4.13 & 4.06 & 4.26\\
 \hline
 $^{76}$Ge & 4.099(1) & 4.06 & 3.99 & 4.16 & 4.10 & 4.30\\
\end{tabular}}
\caption{Proton ($R_p$) and neutron ($R_n$) root-mean-square radii for germanium isotopes. The experimental values of $R_p$ are derived from the measured charge radii reported in Ref.~\cite{Fricke:1995zz}, using the procedure described in Ref.~\cite{Cadeddu:2020lky}. The theoretical values are those obtained from HF+BCS calculations carried out with the D1S~\cite{Berger:1991zza} and D1M~\cite{Chabanat:1997un} interactions, and from NSM predictions~\cite{Hoferichter:2020osn}.}\label{tab:GeRadius}
\end{table}
Here, we consider the values obtained from the recent nuclear shell model (NSM) estimate of the corresponding neutron skins (i.e. the differences between the neutron and the proton rms radii) in Ref.~\cite{Hoferichter:2020osn} (see also Ref.~\cite{PhysRevD.111.033003}). The considered proton and neutron rms radii are summarized in Tab.~\ref{tab:GeRadius}.\\
Given that the differences in the prediction for the radii of different isotopes is below the current experimental precision, when using the Helm phenomenological model, we consider as a reference for our analysis the average values of the nuclear rms proton and neutron radii weighted by the isotopic composition, namely $4.08\,\rm{fm}$ and $4.22\,\rm{fm}$ respectively.
Moreover, we investigate the uncertainties related to the
nuclear model~\cite{Co:2020gwl} by considering also proton and neutron density distributions obtained
by carrying out Hartree-Fock plus Bardeen-Cooper-Schrieffer calculations (HF+BCS)~\cite{Co:2021ijy}. These calculations have been performed by using two  different parameterisations of the 
effective nucleon-nucleon finite-range interaction of Gogny type, namely the D1M~\cite{Chabanat:1997un} and the D1S~\cite{Berger:1991zza} forces. The corresponding rms proton and neutron distribution radii have been reported in Tab.~\ref{tab:GeRadius}.
Therefore, the FF for protons (neutrons) is evaluated as 
\begin{align}\nonumber 
    F_{Z(N)}&= \frac{1}{Z(N)}
    \int d^3r\;e^{i\textbf{q}\cdot\textbf{r}}\rho_{Z(N)}(r)\\
    &=\frac{4\pi}{Z(N)}\int_0^{\infty} dr\;r^2\; \rho_{Z(N)}(r)j_0(qr),
    \label{eq:FF}
\end{align}
where $j_0$ is the zero-order Bessel function. 

\begin{figure}
    \centering
       \topinset{(a)}{
     {\includegraphics[width=0.95\linewidth]{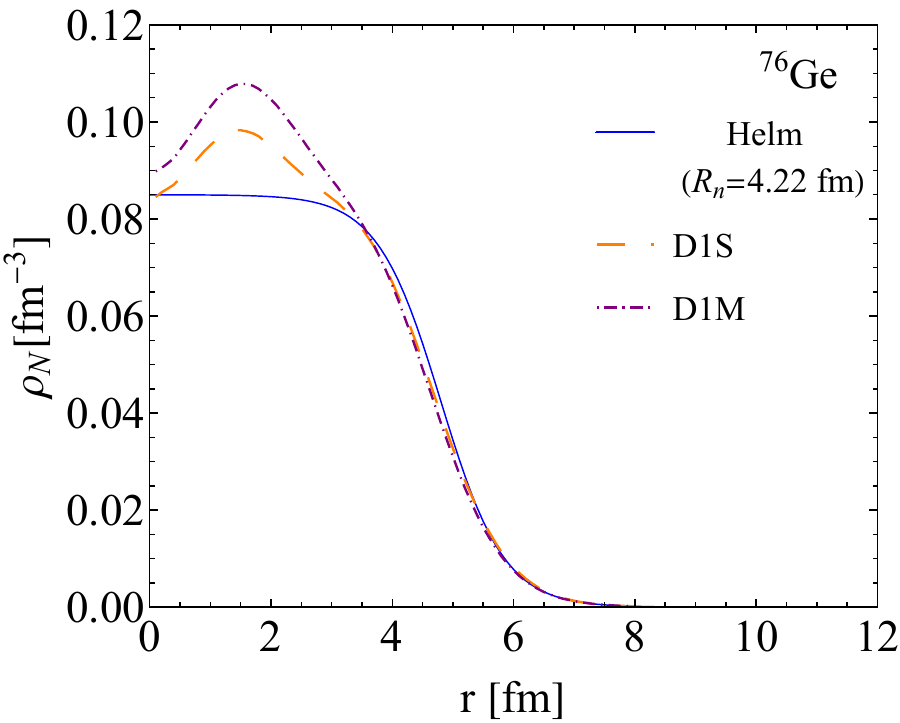}}  
                }{3.5cm}{2cm} 
                \\
                  \topinset{(b)}{
     {\includegraphics[width=0.95\linewidth]{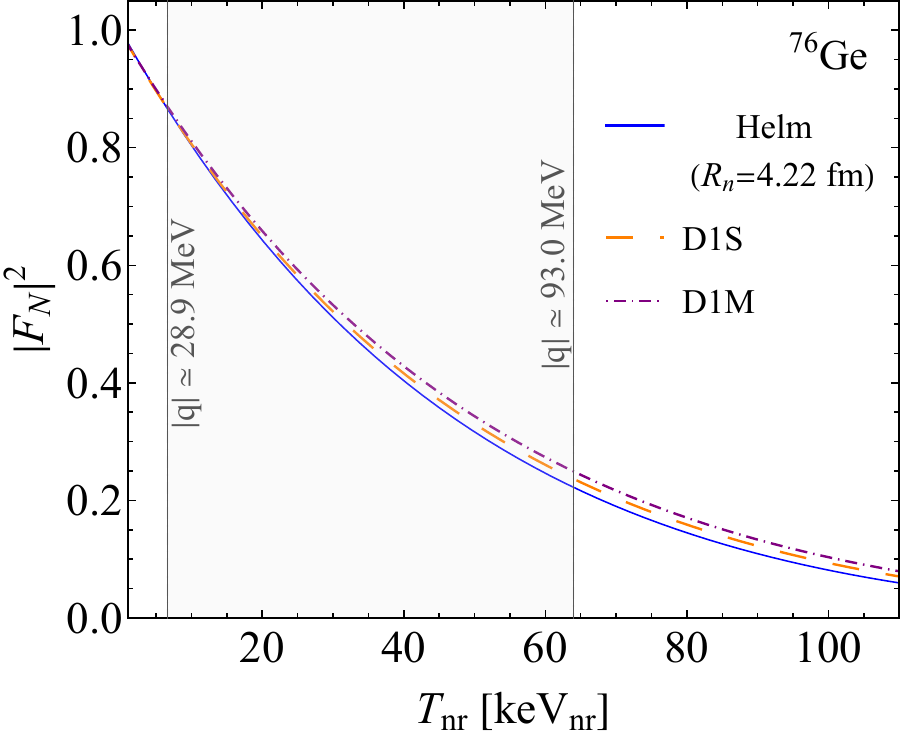}}   
                }{3.5cm}{2cm}
  \caption{(a) Neutron density distributions of the  $^{76}$Ge nucleus obtained with the Helm model 
    by using an average neutron rms radius $R_n^{\rm NSM}=4.22\;\rm fm$ as 
    indicated by the NSM~\cite{Hoferichter:2020osn} (solid blue line) and with HF+BCS calculations 
    obtained by using the
    D1M (dashed-dotted purple) or D1S (dashed orange) Gogny forces. (b) Related FFs, see Eq.~(\ref{eq:FF}), of the    
    neutron densities shown in panel (a). The gray area indicates the region of interest of the COHERENT germanium experiment.}
  \label{fig:nuclearmodels}
\end{figure}

We show in Fig.~\ref{fig:nuclearmodels}~(a) the neutron densities of the $^{76}$Ge nucleus obtained by the HF+BCS calculation
and compare it with our reference Helm neutron distribution obtained by considering the average 
neutron rms radius $R_n^{\rm NSM}=4.22\;\rm fm$.
In Fig.~\ref{fig:nuclearmodels}~(b) we show the related FFs. The sensitivity of the CE$\nu$NS results to the use of the 
different FFs is discussed in the next section.

\section{Anaysis of \cenns data}

We analysed the reactor CONUS+~\cite{Ackermann:2025obx} and TEXONO~\cite{TEXONO:2024vfk} data 
by using the procedure outlined in detail
in Ref.~\cite{AtzoriCorona:2025ygn}. 
We consider, in addition, 
the
data of the recent $\nu$GeN CE$\nu$NS experiment~\cite{nuGeN:2025mla}, which employs a 1.41~kg high-purity, low-threshold germanium detector located at
11.1 m from the Kalinin Nuclear Power Plant. We parametrize the $\nu$GeN reactor neutrino flux using the prescriptions 
of Ref.~\cite{Perisse:2023efm}, which yields a total neutrino flux of $\Phi_{\rm est}\simeq 4.4\times10^{13}\;\rm{\nu/(s\;cm^2)}$. Hereafter, we will refer to the joint analysis of CONUS+, $\nu$GeN and TEXONO data as \texttt{Reactors}.\\

When needed for comparisons, the COHERENT CsI~\cite{Akimov:2021dab} and Ar~\cite{COHERENT:2020iec,COHERENT:2020ybo} data are analysed following the strategy explained in detail in Refs.~\cite{AtzoriCorona:2023ktl,AtzoriCorona:2022moj}. For the COHERENT Ge analysis, 
we instead follow the procedure described 
in Ref.~\cite{COHERENT:2025vuz} and the prescriptions summarised in the following. 
The total neutrino flux from pion decays at rest includes three components: prompt $\nu_\mu$'s from pion decay, and the $\nu_e$ and $\bar{\nu}_\mu$ delayed components, from the subsequent muon decay. The neutrino flux depends on the number of protons-on-target $N_{\mathrm{POT}}=1.96\cdot10^{22}$, the number of neutrinos produced for each POT, $(0.288\pm0.029)$, 
and the baseline between the source and the detector, $L=19.2\;\rm{m}$~\cite{COHERENT:2020ybo}.
In each nuclear-recoil energy-bin $i$, the expected \cenns event number $N^\mathrm{CE \nu NS}_{i}$ on a germanium target is given by
\begin{align}\label{N_cevns}\nonumber
N_{i}^{\mathrm{CE}\nu\mathrm{NS}}
&=
N_T
\int_{T_{\mathrm{nr}}^{i}}^{T_{\mathrm{nr}}^{i+1}}
\hspace{-0.3cm}
d T_{\mathrm{nr}}\,
A(T_{\mathrm{nr}})
\int_{0}^{T^{\prime\text{max}}_{\text{nr}}}
\hspace{-0.3cm}
dT'_{\text{nr}}
\,
R(T_{\text{nr}},T'_{\text{nr}})\\
&\times \int_{E_{\text{min}}(T'_{\text{nr}})}^{E_{\text{max}}}
\hspace{-0.3cm}
d E
\sum_{\nu_{\ell}=\nu_{e}, \nu_{\mu}, \bar{\nu}_{\mu}}
\frac{d N_{\nu_{\ell}}}{d E}
\frac{d \sigma_{\nu_{\ell}-\mathcal{N}}}{d T'_{\mathrm{nr}}}.
\end{align}
Here, $d N_{\nu_{\ell}}/d E$ is the neutrino flux, $T_{\text{nr}}$ is the reconstructed nuclear recoil kinetic energy,
$T'_{\text{nr}}$ is the true nuclear recoil kinetic energy,
$A(T_{\text{nr}})$ is the energy-dependent detector efficiency~\cite{COHERENT:2025vuz},
$R(T_{\rm{nr}},T_{\rm{nr}}')$ is the energy resolution~\cite{COHERENT:2025vuz},
$T^{\prime\text{max}}_{\text{nr}} \simeq 2 E_{\text{max}}^2 / M$,
$E_{\text{max}} = m_\mu/2 \sim 52.8$~MeV,
$E_{\text{min}}(T'_{\text{nr}}) \simeq \sqrt{MT'_\text{nr}/2}$,
$m_\mu$ being the muon mass, and $N_T$ is the number of target atoms in the detector per unit mass. 
It is important to consider that the energy actually observed in the detector is the electron-equivalent recoil energy $T_{e}$, which is transformed into the nuclear recoil energy $T_{\mathrm{nr}}$ in the \cenns rate by inverting the relation
\begin{equation}\label{Qf}
T_{\mathrm{e}}=f_{Q}\left(T_{\mathrm{nr}}\right) T_{\mathrm{nr}}.
\end{equation}
Here $f_Q$ is the quenching factor, whose dependence on the recoil energy is still being actively investigated. A precise understanding of its behaviour is essential for interpreting experimental results~\cite{Colaresi:2022obx,AtzoriCorona:2023ais}, with important consequences for the extraction of SM and beyond parameters~\cite{Li:2025pfw}.
In this work we adopt the Lindhard model~\cite{Lindhard_theo} with $k = 0.162 \pm 0.004$ as measured by the CONUS Collaboration~\cite{Bonhomme:2022lcz}.
Accounting for the time structure of the COHERENT data is crucial for distinguishing between the different neutrino components. We extract the arrival-time distribution of the different neutrino components from Ref.~\cite{M7Germanio}, and incorporate the timing information by dividing the theoretical CE$\nu$NS event numbers $N^{\text{CE}\nu\text{NS}}_i$ in Eq.~\eqref{N_cevns} into time bins of $4\;\mu\text{s}$. This yields $N^{\text{CE}\nu\text{NS}}_{ij}$, where $i$ and $j$ label the energy and time bins, respectively.
To compare the prediction with the data, we perform a simultaneous fit of beam-on (ON) 
and beam-off (OFF) data by using the least-squares function~\cite{COHERENT:2025vuz}
\begin{equation}\label{eq:chi2tot}
\chi^2 =  
\chi^2_{\text{ON}} (\eta, \beta) + \chi^2_{\text{OFF}} (\beta) 
+ \left( \frac{\eta-1}{\sigma_{\eta}} \right)^2+ \left( \frac{\beta-1}{\sigma_{\beta}} \right)^2,
\end{equation}
where
\begin{align}\nonumber
\chi^2_{\text{ON}} &= 2 \sum_{\substack{i=1 \\ i \neq \{4,5\}}}^{9} \sum_{j=1}^{10}  \bigg( 
\eta N^{\text{CE}\nu\text{NS}}_{ij} 
+ \beta N^{\text{SSB}}_{ij} 
- N^{\text{exp, ON}}_{ij} \\
&\quad + N_{ij}^{\text{exp, ON}} 
\mathrm{ln} \left[ \frac{N_{ij}^{\text{exp, ON}}}
{\eta N^{\text{CE}\nu\text{NS}}_{ij} 
+ \beta N^{\text{SSB}}_{ij} } \right] \bigg),
\end{align}
and
\begin{equation}
\begin{aligned}
\chi^2_{\text{OFF}} &= 2 \sum_{\substack{i=1 \\ i \neq \{4,5\}}}^{9} \sum_{j=1}^{10} \bigg( 
\beta N^{\text{SSB}}_{ij} 
- N^{\text{exp, OFF}}_{ij} \\
&\quad + N_{ij}^{\text{exp, OFF}} 
\mathrm{ln} \left[ \frac{N_{ij}^{\text{exp, OFF}}}
{\beta N^{\text{SSB}}_{ij} } \right] \bigg).
\end{aligned}
\end{equation}
Here, we are following the prescription from the COHERENT collaboration to remove the energy bins in the range $[8.5,11.5]\;\rm keV_{ee}$\footnote{
It is important to highlight that due the 2D binning of the data provided by the COHERENT collaboration, we are excluding the energy range $[7.5,11.5]\;\rm keV_{ee}$, while the official result excludes the energy range $[8.5,11.5]\;\rm keV_{ee}$.}, which corresponds to $i=4,5$~\cite{COHERENT:2025vuz}.
Moreover, $N_{i,j}^{\rm exp, ON\;(OFF)}$ is the observed number of events from beam-ON (OFF) data as extracted from Ref.~\cite{COHERENT:2025vuz} in the $i-$th (energy) and $j-$th (time) bin, while $N_{ij}^{\rm SSB}$ is the steady state background (SSB) from internally-triggered data. The latter is extracted from Fig.~3 of Ref.~\cite{COHERENT:2025vuz},
where the spectral shape of the SSB is provided in a $40\;\mu s$ time window, for the exposure of the beam-ON data set. The two-dimensional distribution of the SSB is obtained by assuming it to be constant over time. Finally, $N_{i,j}^{\rm CE\nu NS}$ is the predicted number of CE$\nu$NS events that depends on the physics model under consideration.
The systematic uncertainty on the signal prediction is $\sigma_\eta=10.3\%$ and it is dominated by the uncertainty on the neutrino flux (10\%), but includes also the uncertainty on the detector location (0.5\%), energy calibration (1\%), active mass (1\%) and form factors (1\%). The systematic uncertainty on the SSB is $\sigma_\beta=1\%$, thanks to the fact that the simultaneous fit of ON and OFF data allows one to constrain the normalisation of the SSB in the fit precisely. \\
\begin{figure}[t]
    \includegraphics[width=0.8\linewidth]{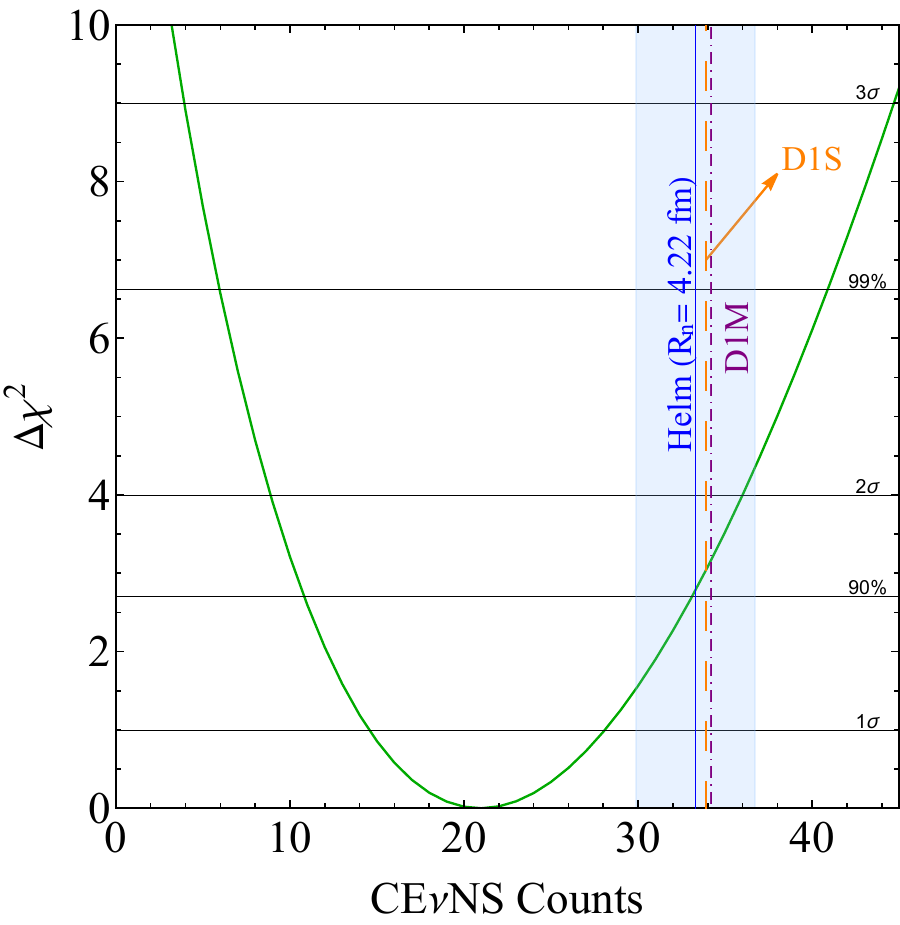}
    \caption{Marginal $\Delta\chi^2$ profile as a function of the number of \cenns events. For comparison, the \cevns prediction for different nuclear models is shown by the vertical lines: using a Helm neutron FF with an average neutron rms radius $R_n^{\rm NSM}=4.22\;\rm fm$ as predicted by NSM~\cite{Hoferichter:2020osn} (solid blue), or parameterizing the nuclear density distribution with a D1M (dashed-dotted purple) or D1S (dashed orange) Gogny type model. The light blue area indicates the systematic uncertainty on the theoretical prediction $\sigma_\eta=10.3\%$.}
    \label{fig:Significance}
\end{figure}

We employ Eq.~(\ref{eq:chi2tot}) to extract the experimental number of \cevns events obtained leaving the \cevns normalisation free to vary in the fit, while removing the systematic uncertainty $\sigma_\eta$. 
The marginal $\Delta\chi^2=\chi^2-\chi^2_{\rm{min}}$ distribution is shown in Fig.~\ref{fig:Significance}, which is in excellent agreement with that released by the COHERENT collaboration~\cite{COHERENT:2025vuz}.
We find about $21.0^{+7.1}_{-6.4}$ \cenns events, which should be compared with our theoretical prediction of 33.3$\pm$3.4 events assuming a Helm neutron FF with an average neutron rms radius $R_n^{\rm NSM}=4.22\;\rm fm$ from NSM, which increases to 33.9$\pm$3.5 (34.2$\pm$3.5) considering the D1S (D1M) model for the neutron nuclear density. Our theoretical prediction obtained using the NSM is 1.7$\sigma$ above the best-fit value. This small tension increases to about 1.8$\sigma$ when assuming the D1S and D1M nuclear models.\\
\begin{figure}[t]
    \includegraphics[width=0.8\linewidth]{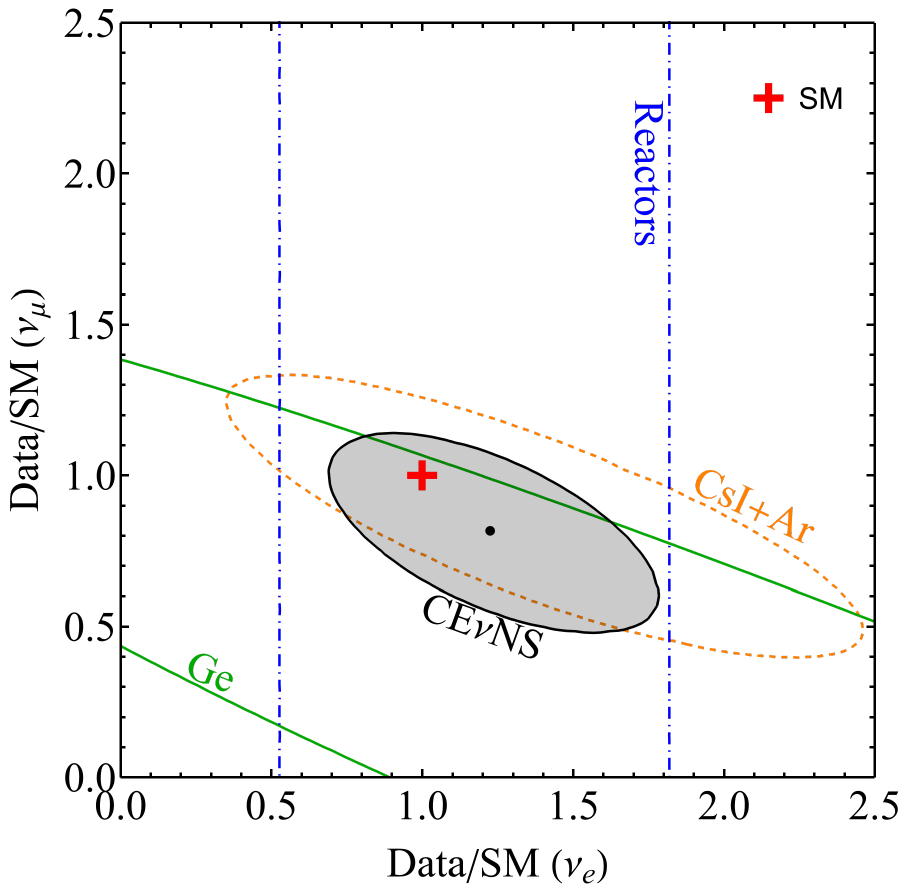}
    \caption{Agreement between \cenns data and SM predictions for COHERENT CsI+Ar (orange dashed contour), reactors (blue dotted-dashed band), COHERENT germanium (green solid contour) and the combined analysis (black region), separate for electron and muon neutrinos. Contours are shown at 90\% CL, and the black point represents the best-fit, while the red cross indicates perfect agreement with the SM prediction.}
    \label{fig:Norm2D}
\end{figure}
In order to search for potential flavour-dependent effects in the COHERENT germanium data and exploit the power of the timing information, we split the signal normalisation, $\eta=\rm Data/SM$, according to the neutrino flavour. This strategy, introduced in Ref.~\cite{AtzoriCorona:2025ygn}, allows for a meaningful comparison with reactor \cenns data, which are sensitive only to the $\nu_e$ flavour. The result is shown in Fig.~\ref{fig:Norm2D} at the 90\% confidence level (CL). Within this combined analysis, we find a general agreement among all \cenns probes, obtaining
\begin{align}\nonumber
    &\textrm{Data/SM}\;(\nu_e)=1.24^{+0.25}_{-0.26}(1\sigma),^{+0.42}_{-0.43}(90\%),^{+0.51}_{-0.52}(2\sigma),\\
    &\textrm{Data/SM}\;(\nu_\mu)=0.78^{+0.17}_{-0.14}(1\sigma),^{+0.27}_{-0.24}(90\%),^{+0.33}_{-0.29}(2\sigma).
\end{align}
The best fit indicates a preference for a slightly higher normalisation of the electronic flavour, but consistent with the SM 
within 1$\sigma$, while a slight under fluctuation is observed for the muonic flavour, at the 1$\sigma$ level. 
Nevertheless, both normalisations remain consistent with one at the 90\%~CL. 
Future datasets with improved statistical precision will allow for a more accurate testing of SM predictions.

\section{Implications for Electroweak and Nuclear Physics}

\begin{figure}[t]
    \includegraphics[width=0.8\linewidth]{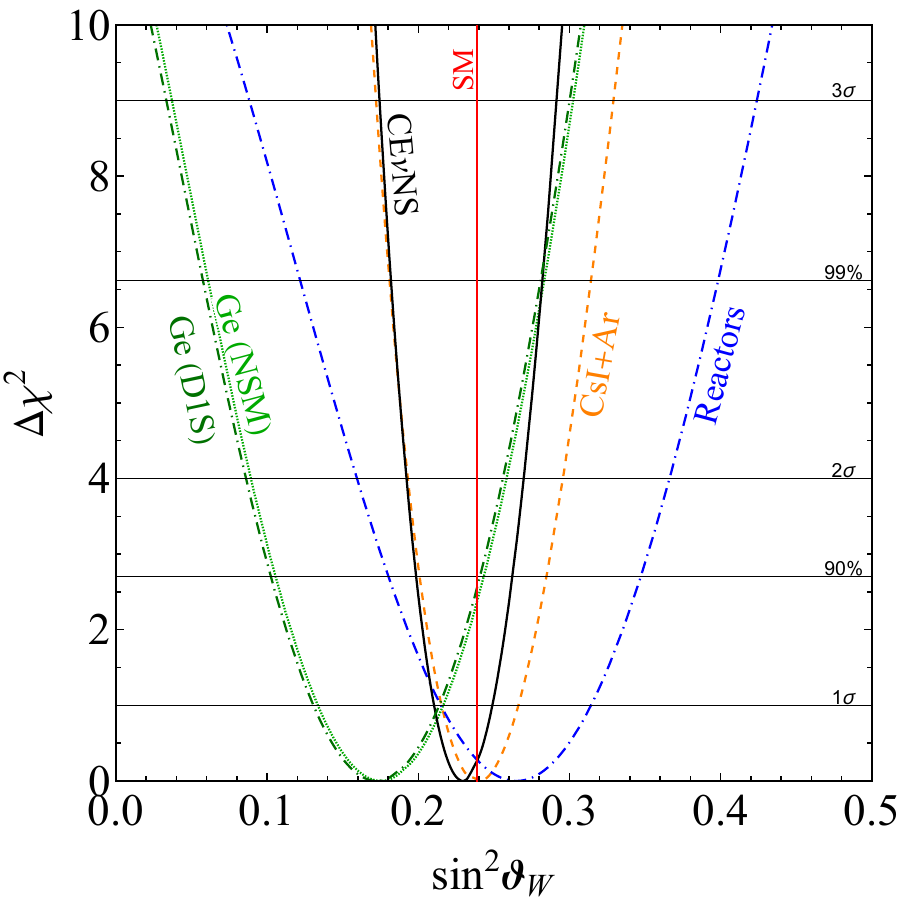}
    \caption{Constraints on the weak mixing angle at different confidence levels from the joint analysis of CsI and Ar data (dashed orange), reactors (dashed-dotted blu) and COHERENT germanium (green), the latter assuming two different parameterisations for the nuclear structure. The combined analysis is shown in black while the SM prediction is shown by the red vertical line.
    }
    \label{fig:wma1D}
\end{figure}

In Fig.~\ref{fig:wma1D} we show the constraints on the value of the weak mixing angle, $\sin^2\vartheta_W$, 
from the COHERENT Ge data-set, by fixing the neutron nuclear radius to the theoretical predictions from NSM and the D1S model (which practically overlaps with the D1M one). We notice that the extracted values 
of the weak mixing angle are not strongly dependent on the parameterisation of the nuclear structure, making \cenns a robust probe for studying SM key parameters.
In the same figure, we also show the constraints from the joint CsI~\cite{AtzoriCorona:2023ktl} and Ar~\cite{Cadeddu:2020lky} analysis, from reactor data, i.e. CONUS+, TEXONO with the inclusion of $\nu$GeN, and the result of a combined analysis when fixing the neutron nuclear radius to the NSM prediction. 
The result of this combined analysis is
\begin{align}
\sin^2\theta_W=0.229^{+0.020}_{-0.019}\,(1\sigma), ^{+0.040}_{-0.037}\,(2\sigma),\pm0.06\;(3\sigma).
\label{eq:sinw}
\end{align}
This result represents the state-of-the-art determination on the low-energy value of the weak mixing angle from \cenns probes, and it is consistent with the SM and the recent re-evaluation in Ref.~\cite{AtzoriCorona:2024vhj}.\\

\begin{figure}[t]
    \includegraphics[width=0.9\linewidth]{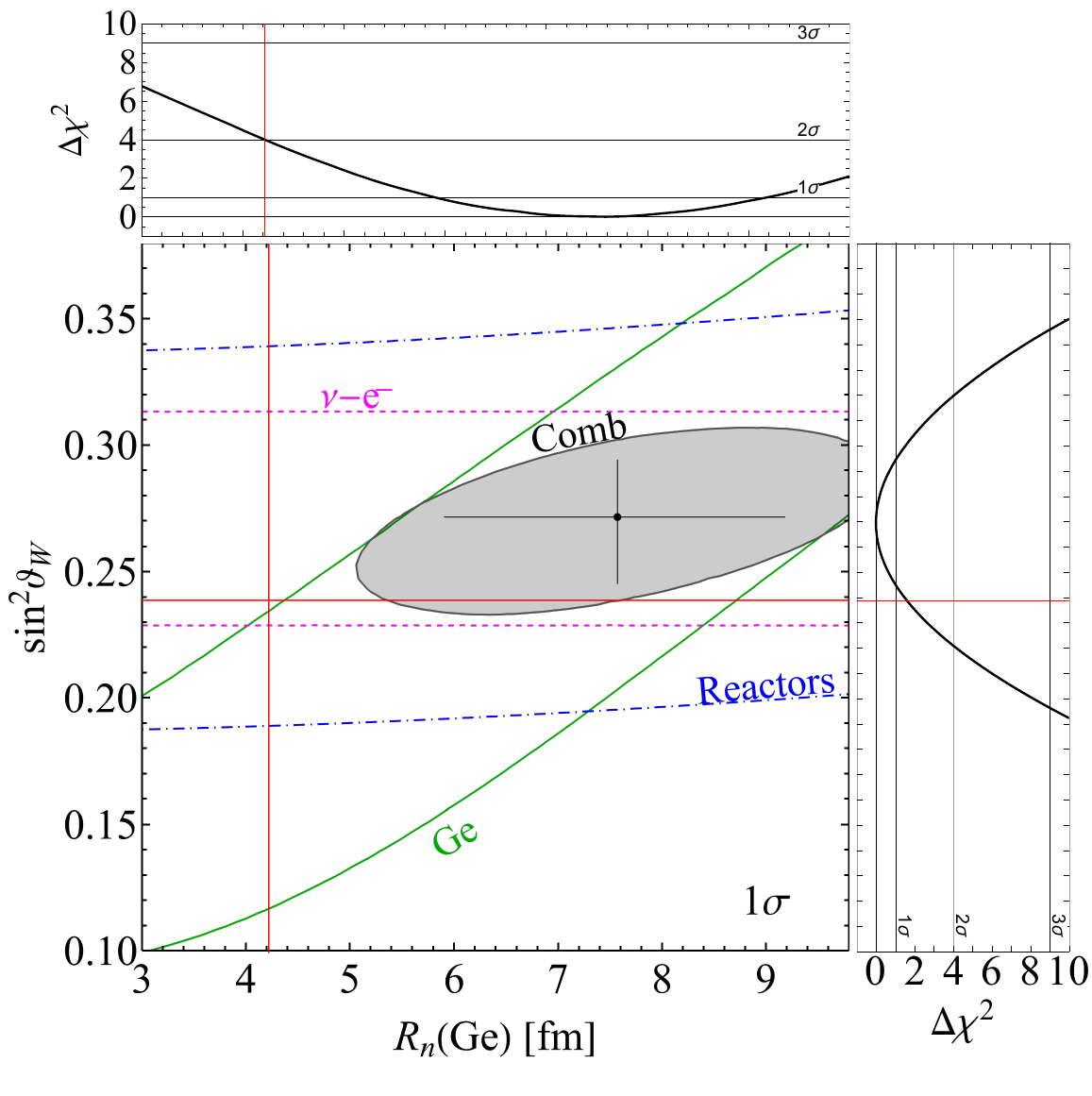}
    \caption{Constraints obtained by fitting the weak mixing angle and the average rms Ge neutron radius on COHERENT Ge data (solid green band) and reactors (dashed-dotted blue band), as well as the prior on the weak mixing angle from $\nu-e^-$ data (dashed magenta band) and the result of a combined analyis (black contour) at 1$\sigma$ CL. The upper (right) panel shows the one-dimensional marginalisation on neutron radius (weak mixing angle) from the combined analysis at different CLs. The red lines indicate the SM low-energy value of the weak mixing angle and the NSM prediction for $R_n(\rm{Ge})$.}
    \label{fig:2DCOH}
\end{figure}

When using electroweak probes with nuclei, the weak mixing angle is known to be correlated with the neutron nuclear radius~\cite{Cadeddu:2021ijh, Cadeddu:2024baq, AtzoriCorona:2024vhj, Cadeddu:2018izq, AtzoriCorona:2023ktl}. 
Therefore, we also perform a two-dimensional analysis
by considering both $\sin^2\vartheta_W$ and $R_n(\rm{Ge})$ as free parameters. The outcome is shown in Fig.~\ref{fig:2DCOH}, where we also display the result of a fit to reactor germanium data under this configuration. This result clearly highlights the complementarity between the findings obtained by fitting data from accelerator-based neutrino experiments and those from reactor-based experiments. The former exhibits a band with a strong correlation between the weak mixing angle and the neutron nuclear radius.
On the contrary the reactor \cenns data are practically insensitive to $R_n(\rm{Ge})$. 
To further restrict the phase-space available on the weak mixing angle, in our analysis we also consider low-energy $\nu-e^-$ scattering data from TEXONO~\cite{TEXONO:2009knm}, LSND~\cite{LSND:2001akn}, LAMPF~\cite{Allen:1992qe}, LZ~\cite{LZ:2022lsv}, PandaX~\cite{PandaX:2022ood} and XENONnT~\cite{XENON:2022ltv} electron recoil data, as described in Ref.~\cite{AtzoriCorona:2025xwr}. 
By using all these data we obtain a clean low-energy determination of the weak mixing angle, which is independent of the neutron nuclear radius.
By combining all these inputs we find, at the $1\sigma$ CL, the following values
\begin{equation}
    \sin^2\vartheta_W=0.271^{+0.023}_{-0.026};\;\ R_n(\rm{Ge})=7.6^{+1.6}_{-1.7}\;\rm{fm}.
    \label{eq:2D_BF}
\end{equation}
The value of the nuclear neutron radius emerging from this analysis
is exceptionally high with respect to those obtained in the HF+BCS calculations
and excludes the NSM prediction~\cite{Hoferichter:2020osn} at about $2\sigma$. This disagreement is driven by the positive correlation of $R_n(\rm{Ge})$ and $\sin^2\vartheta_W$ in COHERENT data. In addition, we remark that
the $\nu-e^-$ scattering data slightly prefer higher 
$\sin^2\vartheta_W$ values than those indicated by the SM, and this
selects a region of the parameter space in disagreement with expectations. 
The adoption of the best-fit values in Eq.~(\ref{eq:2D_BF}) 
improves the fit to COHERENT Ge data by $\Delta\chi^2\simeq-2.7$, 
when compared to the use of the expected SM and NSM $\sin^2\vartheta_W$ and $R_n(\rm{Ge})$ values, respectively. This can be physically explained by the fact that a larger nuclear radius is responsible for an enhanced suppression in the cross section, which makes the prediction more in agreement with the under-fluctuations of events observed in the COHERENT Ge data.

We also evaluated the constraints on the average rms germanium nuclear radius when fixing the weak mixing angle to its SM value, namely $0.23873\pm0.00005$~\cite{PhysRevD.110.030001},
and we obtain
\begin{align}\nonumber
    R_n(\textrm{Ge})&=6.5\pm1.4\,(1\sigma), \pm3.0\,(2\sigma),\\
    &<11.8\,(3\sigma)\;\rm{fm}.
    \label{eq:Rn}
\end{align}
It is worth highlighting that the fact that the COHERENT collaboration removed the energy bins in the range $[7.5,11.5]\;\rm{keV_{ee}}$, which corresponds to $[27.5,39.0]\;\rm{keV_{nr}}$, is not beneficial for accurate studies of the nuclear structure, as a crucial part of the spectral shape information is lost. In fact, in this energy regime, the nuclear neutron form factor $|\rm F_N|^2$ varies from $[0.56,0.44]$ when assuming the nuclear radius from nuclear shell models, as shown in Fig.~\ref{fig:nuclearmodels}~(b). Moreover, a \cenns analysis in the full energy spectrum would not only be relevant to reduce the uncertainty on $R_n(\rm{Ge})$, but would also be crucial to search for insights beyond the nuclear theory.  \\

\begin{figure}[t]
    \includegraphics[width=0.8\linewidth]{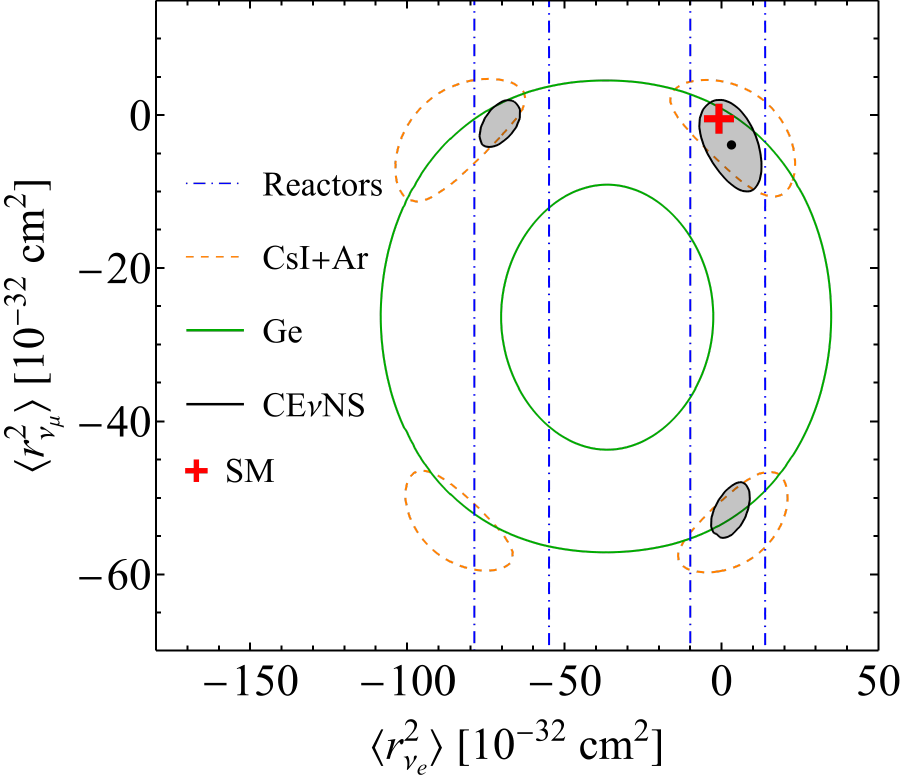}
    \caption{Allowed regions at 90\% CL on the electronic and muonic neutrino charge radii considering a momentum-dependence in the neutrino charge radii radiative corrections~\cite{AtzoriCorona:2024rtv}, from the analysis COHERENT germanium, COHERENT CsI and Ar, reactor CE$\nu$NS data, together with their combined analysis. The red cross indicates the SM values reported in Eqs.~(\ref{eq:cr-e}) and (\ref{eq:cr-mu}), while the black dot represents the best-fit of the combined analysis.}
    \label{fig:NCR}
\end{figure}
Another fundamental SM parameter which can be constrained by \cenns processes
is the neutrino charge radius. Here, we also account for the momentum-dependence 
of the neutrino charge-radius radiative correction, as described in Ref.~\cite{AtzoriCorona:2024rtv}. 
The results of the fit to all available \cenns data are summarised in Fig.~\ref{fig:NCR} 
by using a $\langle{r_{\nu_{e}}^{2}}\rangle$ and $\langle{r_{\nu_{\mu }}^{2}}\rangle$ plane. 
Due to the limited statistics, at 90\% CL COHERENT Ge displays a full degeneracy among these two parameters which can be broken thanks to the complementarity with reactor data which are only sensitive to the electronic neutrino flavour.
Intriguingly, even if the 2D contours in Fig.~\ref{fig:NCR} present three degenerate regions at the 90\% CL, when considering 2 degrees of freedom ($\Delta\chi^2=4.6$), the corresponding marginal 1D distributions select the SM region at the 1$\sigma$ CL for both flavors, namely
\begin{align}
    \langle{r_{\nu_{e}}^{2}}\rangle=&\left(3.2\pm4.7\right)\times10^{-32}\, \mathrm{cm}^2,\\
    \langle{r_{\nu_{\mu }}^{2}}\rangle\,=&-3.9^{+2.8}_{-2.7}\times10^{-32}\; \mathrm{cm}^2 .
    \label{eq:CR_limit}
\end{align}
\section{Interpretations and conclusions}

We investigate the possible sources of the small discrepancy between theoretical predictions and data of the COHERENT germanium dataset. Nuclear physics does not appear to resolve this issue, as the best-fit value for the neutron rms radius is found to be unphysically large, see Eq.~(\ref{eq:2D_BF}). Invoking physics Beyond the Standard Model
(BSM) is also challenging in this case, as most BSM scenarios, such as neutrino magnetic moments, millicharges, or light mediator bosons, typically predict an excess of events compared to the SM, whereas the germanium data shows a deficit.

\begin{figure}[t]
    \includegraphics[width=0.84\linewidth]{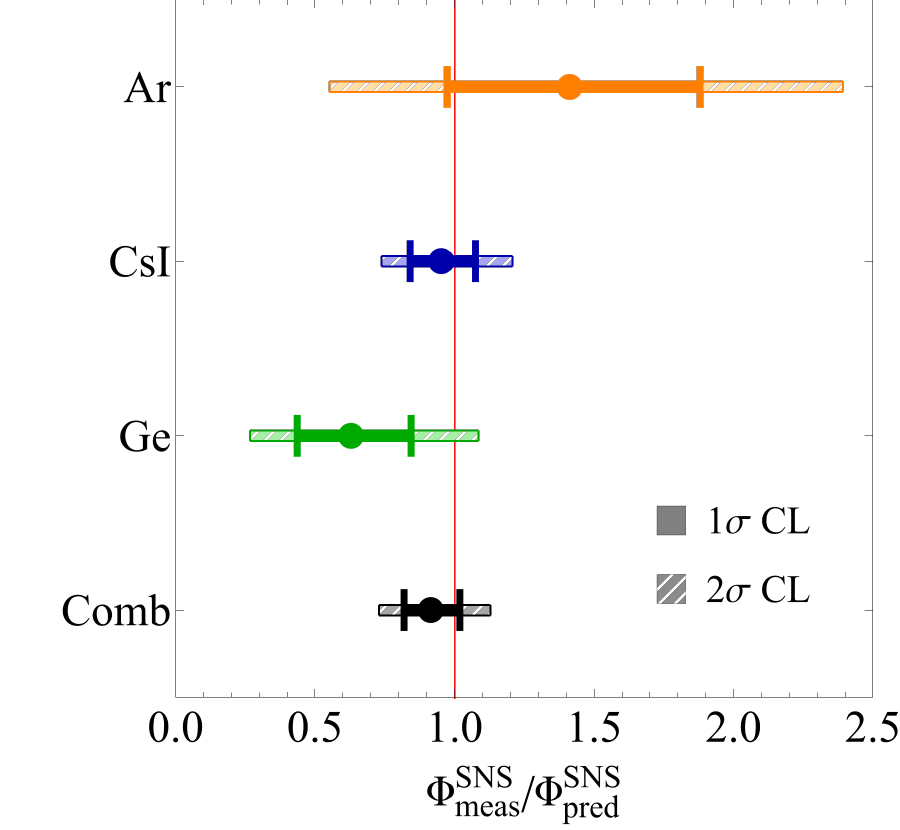}
    \caption{
    Measured neutrino flux at the SNS relative to the total expected neutrino flux during the full data-taking period for Ar, CsI, and Ge data and their combination.}
    \label{fig:FLUXCOH}
\end{figure}
We therefore focus on a possible systematic effect related to the neutrino flux measured at the SNS, $\Phi^{\rm{SNS}}_{\rm{meas}}$, which represents the dominant uncertainty in the \cenns signal. To explore this, we show in Fig.~\ref{fig:FLUXCOH} the constraints obtained by allowing the normalisation of the SNS neutrino flux to vary freely in the fit. This normalisation is taken relative to the total predicted neutrino flux, $\Phi^{\rm{SNS}}_{\rm{pred}}$, during the data-taking period. We obtain
\begin{align}
\textrm{Ar:}\quad\frac{\Phi^{\rm{SNS}}_{\rm{meas}}}{\Phi^{\rm{SNS}}_{\rm{pred}}}&=1.4^{+0.5}_{-0.4}\,(1\sigma),^{+1.0}_{-0.9}\,(2\sigma),\\\textrm{CsI:}\quad\frac{\Phi^{\rm{SNS}}_{\rm{meas}}}{\Phi^{\rm{SNS}}_{\rm{pred}}}&=0.95^{+0.12}_{-0.11}\,(1\sigma),^{+0.25}_{-0.21}\,(2\sigma),\\
\textrm{Ge:}\quad\frac{\Phi^{\rm{SNS}}_{\rm{meas}}}{\Phi^{\rm{SNS}}_{\rm{pred}}}&=0.63^{+0.22}_{-0.19}\,(1\sigma),^{+0.46}_{-0.36}\,(2\sigma),\\
\textrm{Comb:}\quad\frac{\Phi^{\rm{SNS}}_{\rm{meas}}}{\Phi^{\rm{SNS}}_{\rm{pred}}}&=0.92\pm0.10\,(1\sigma),^{+0.21}_{-0.19}\,(2\sigma). \label{eq:combflux}
\end{align}

The combined result in Eq.~(\ref{eq:combflux}) indicates a possible flux deficit, with a best-fit combined normalisation factor of 0.92. However, this value remains compatible with unity at the 1$\sigma$ level and falls within the 10\% systematic uncertainty quoted by the collaboration. This downward fluctuation is primarily driven by the CsI and Ge datasets, while the Ar dataset shows a slight upward fluctuation, albeit with a much larger uncertainty.

Although this analysis does not provide conclusive evidence for a flux-related systematic effect that could explain the event deficit observed in COHERENT Ge, it clearly highlights the importance of reducing the associated uncertainty. 
For this purpose, COHERENT is developing a new detector~\cite{COHERENT:2021xhx} based on heavy water $\rm D_2 O$, which aims to reduce the neutrino flux uncertainty to the 2–3\% level. The detector will be positioned approximately 20 meters from the SNS target and is expected to achieve better than 5\% statistical uncertainty on the neutrino flux within two years of operation. This heavy-water detector will also serve as the first module of a two-module system, ultimately designed to provide high-precision flux measurements. 
This will significantly improve the precision on both SM and BSM observables and help clarify whether the tension in the COHERENT Ge dataset is merely a statistical fluctuation or an indication of new physics.\\

Despite the small underfluctuation observed in the COHERENT germanium data, our combined analysis of all available CE$\nu$NS measurements yields a consistent and robust global picture. This joint fit enables precise determinations of the weak mixing angle and the neutrino charge radii, demonstrating the value of integrating diverse experimental inputs.
In particular, the complementarity between accelerator- and reactor-based CE$\nu$NS experiments 
is essential to break parameter degeneracies and to reduce uncertainties.
This unified approach not only maximises the scientific return of current data but also lays solid foundations for future precision CE$\nu$NS studies. As the field advances toward next-generation detectors with improved sensitivity, varied target materials, and broader physics goals, such global strategies will be essential to fully exploit the potential of \cenns.

\begin{acknowledgements}
The work of C. Giunti is partially supported by the PRIN 2022 research grant Number 2022F2843L funded by MIUR.
\end{acknowledgements}

\bibliography{ref}

\begin{thebibliography}{83}%
\makeatletter
\providecommand \@ifxundefined [1]{%
 \@ifx{#1\undefined}
}%
\providecommand \@ifnum [1]{%
 \ifnum #1\expandafter \@firstoftwo
 \else \expandafter \@secondoftwo
 \fi
}%
\providecommand \@ifx [1]{%
 \ifx #1\expandafter \@firstoftwo
 \else \expandafter \@secondoftwo
 \fi
}%
\providecommand \natexlab [1]{#1}%
\providecommand \enquote  [1]{``#1''}%
\providecommand \bibnamefont  [1]{#1}%
\providecommand \bibfnamefont [1]{#1}%
\providecommand \citenamefont [1]{#1}%
\providecommand \href@noop [0]{\@secondoftwo}%
\providecommand \href [0]{\begingroup \@sanitize@url \@href}%
\providecommand \@href[1]{\@@startlink{#1}\@@href}%
\providecommand \@@href[1]{\endgroup#1\@@endlink}%
\providecommand \@sanitize@url [0]{\catcode `\\12\catcode `\$12\catcode `\&12\catcode `\#12\catcode `\^12\catcode `\_12\catcode `\%12\relax}%
\providecommand \@@startlink[1]{}%
\providecommand \@@endlink[0]{}%
\providecommand \url  [0]{\begingroup\@sanitize@url \@url }%
\providecommand \@url [1]{\endgroup\@href {#1}{\urlprefix }}%
\providecommand \urlprefix  [0]{URL }%
\providecommand \Eprint [0]{\href }%
\providecommand \doibase [0]{http://dx.doi.org/}%
\providecommand \selectlanguage [0]{\@gobble}%
\providecommand \bibinfo  [0]{\@secondoftwo}%
\providecommand \bibfield  [0]{\@secondoftwo}%
\providecommand \translation [1]{[#1]}%
\providecommand \BibitemOpen [0]{}%
\providecommand \bibitemStop [0]{}%
\providecommand \bibitemNoStop [0]{.\EOS\space}%
\providecommand \EOS [0]{\spacefactor3000\relax}%
\providecommand \BibitemShut  [1]{\csname bibitem#1\endcsname}%
\let\auto@bib@innerbib\@empty
%</preamble>
\bibitem [{\citenamefont {Freedman}(1974)}]{Freedman:1973yd}%
  \BibitemOpen
  \bibfield  {author} {\bibinfo {author} {\bibfnamefont {D.~Z.}\ \bibnamefont {Freedman}},\ }\href {\doibase 10.1103/PhysRevD.9.1389} {\bibfield  {journal} {\bibinfo  {journal} {Phys. Rev. D}\ }\textbf {\bibinfo {volume} {9}},\ \bibinfo {pages} {1389} (\bibinfo {year} {1974})}\BibitemShut {NoStop}%
\bibitem [{\citenamefont {Cadeddu}\ \emph {et~al.}(2023)\citenamefont {Cadeddu}, \citenamefont {Dordei},\ and\ \citenamefont {Giunti}}]{Cadeddu:2023tkp}%
  \BibitemOpen
  \bibfield  {author} {\bibinfo {author} {\bibfnamefont {M.}~\bibnamefont {Cadeddu}}, \bibinfo {author} {\bibfnamefont {F.}~\bibnamefont {Dordei}}, \ and\ \bibinfo {author} {\bibfnamefont {C.}~\bibnamefont {Giunti}},\ }\href {\doibase 10.1209/0295-5075/ace7f0} {\bibfield  {journal} {\bibinfo  {journal} {EPL}\ }\textbf {\bibinfo {volume} {143}},\ \bibinfo {pages} {34001} (\bibinfo {year} {2023})},\ \Eprint {http://arxiv.org/abs/2307.08842} {arXiv:2307.08842 [hep-ph]} \BibitemShut {NoStop}%
\bibitem [{\citenamefont {Atzori~Corona}\ \emph {et~al.}(2025{\natexlab{a}})\citenamefont {Atzori~Corona}, \citenamefont {Cadeddu}, \citenamefont {Cargioli}, \citenamefont {Dordei}, \citenamefont {Giunti},\ and\ \citenamefont {Ternes}}]{AtzoriCorona:2025xwr}%
  \BibitemOpen
  \bibfield  {author} {\bibinfo {author} {\bibfnamefont {M.}~\bibnamefont {Atzori~Corona}}, \bibinfo {author} {\bibfnamefont {M.}~\bibnamefont {Cadeddu}}, \bibinfo {author} {\bibfnamefont {N.}~\bibnamefont {Cargioli}}, \bibinfo {author} {\bibfnamefont {F.}~\bibnamefont {Dordei}}, \bibinfo {author} {\bibfnamefont {C.}~\bibnamefont {Giunti}}, \ and\ \bibinfo {author} {\bibfnamefont {C.~A.}\ \bibnamefont {Ternes}},\ }\href@noop {} {\enquote {\bibinfo {title} {{The Standard Model tested with neutrinos}},}\ } (\bibinfo {year} {2025}{\natexlab{a}}),\ \Eprint {http://arxiv.org/abs/2504.05272} {arXiv:2504.05272 [hep-ph]} \BibitemShut {NoStop}%
\bibitem [{\citenamefont {Akimov}\ \emph {et~al.}(2017)\citenamefont {Akimov} \emph {et~al.}}]{COHERENT:2017ipa}%
  \BibitemOpen
  \bibfield  {author} {\bibinfo {author} {\bibfnamefont {D.}~\bibnamefont {Akimov}} \emph {et~al.} (\bibinfo {collaboration} {COHERENT}),\ }\href {\doibase 10.1126/science.aao0990} {\bibfield  {journal} {\bibinfo  {journal} {Science}\ }\textbf {\bibinfo {volume} {357}},\ \bibinfo {pages} {1123} (\bibinfo {year} {2017})},\ \Eprint {http://arxiv.org/abs/1708.01294} {arXiv:1708.01294 [nucl-ex]} \BibitemShut {NoStop}%
\bibitem [{\citenamefont {Akimov}\ \emph {et~al.}(2018)\citenamefont {Akimov} \emph {et~al.}}]{COHERENT:2018imc}%
  \BibitemOpen
  \bibfield  {author} {\bibinfo {author} {\bibfnamefont {D.}~\bibnamefont {Akimov}} \emph {et~al.} (\bibinfo {collaboration} {COHERENT}),\ }\href {\doibase 10.5281/zenodo.1228631} {\enquote {\bibinfo {title} {{COHERENT Collaboration data release from the first observation of coherent elastic neutrino-nucleus scattering}},}\ } (\bibinfo {year} {2018}),\ \Eprint {http://arxiv.org/abs/1804.09459} {arXiv:1804.09459 [nucl-ex]} \BibitemShut {NoStop}%
\bibitem [{\citenamefont {Akimov}\ \emph {et~al.}(2022)\citenamefont {Akimov} \emph {et~al.}}]{COHERENT:2021xmm}%
  \BibitemOpen
  \bibfield  {author} {\bibinfo {author} {\bibfnamefont {D.}~\bibnamefont {Akimov}} \emph {et~al.} (\bibinfo {collaboration} {COHERENT}),\ }\href {\doibase 10.1103/PhysRevLett.129.081801} {\bibfield  {journal} {\bibinfo  {journal} {Phys. Rev. Lett.}\ }\textbf {\bibinfo {volume} {129}},\ \bibinfo {pages} {081801} (\bibinfo {year} {2022})},\ \Eprint {http://arxiv.org/abs/2110.07730} {arXiv:2110.07730 [hep-ex]} \BibitemShut {NoStop}%
\bibitem [{\citenamefont {Akimov}\ \emph {et~al.}(2021{\natexlab{a}})\citenamefont {Akimov} \emph {et~al.}}]{COHERENT:2020iec}%
  \BibitemOpen
  \bibfield  {author} {\bibinfo {author} {\bibfnamefont {D.}~\bibnamefont {Akimov}} \emph {et~al.} (\bibinfo {collaboration} {COHERENT}),\ }\href {\doibase 10.1103/PhysRevLett.126.012002} {\bibfield  {journal} {\bibinfo  {journal} {Phys. Rev. Lett.}\ }\textbf {\bibinfo {volume} {126}},\ \bibinfo {pages} {012002} (\bibinfo {year} {2021}{\natexlab{a}})},\ \Eprint {http://arxiv.org/abs/2003.10630} {arXiv:2003.10630 [nucl-ex]} \BibitemShut {NoStop}%
\bibitem [{\citenamefont {De~Romeri}\ \emph {et~al.}(2024{\natexlab{a}})\citenamefont {De~Romeri}, \citenamefont {Papoulias}, \citenamefont {Sanchez~Garcia}, \citenamefont {Ternes},\ and\ \citenamefont {T\'ortola}}]{DeRomeri:2024hvc}%
  \BibitemOpen
  \bibfield  {author} {\bibinfo {author} {\bibfnamefont {V.}~\bibnamefont {De~Romeri}}, \bibinfo {author} {\bibfnamefont {D.~K.}\ \bibnamefont {Papoulias}}, \bibinfo {author} {\bibfnamefont {G.}~\bibnamefont {Sanchez~Garcia}}, \bibinfo {author} {\bibfnamefont {C.~A.}\ \bibnamefont {Ternes}}, \ and\ \bibinfo {author} {\bibfnamefont {M.}~\bibnamefont {T\'ortola}},\ }\href@noop {} {\enquote {\bibinfo {title} {{Neutrino electromagnetic properties and sterile dipole portal in light of the first solar CE$\nu$NS data}},}\ } (\bibinfo {year} {2024}{\natexlab{a}}),\ \Eprint {http://arxiv.org/abs/2412.14991} {arXiv:2412.14991 [hep-ph]} \BibitemShut {NoStop}%
\bibitem [{\citenamefont {Akimov}\ \emph {et~al.}(2024)\citenamefont {Akimov} \emph {et~al.}}]{Akimov:2024lnl}%
  \BibitemOpen
  \bibfield  {author} {\bibinfo {author} {\bibfnamefont {D.~Y.}\ \bibnamefont {Akimov}} \emph {et~al.},\ }\href@noop {} {\enquote {\bibinfo {title} {{First constraints on the coherent elastic scattering of reactor antineutrinos off xenon nuclei}},}\ } (\bibinfo {year} {2024}),\ \Eprint {http://arxiv.org/abs/2411.18641} {arXiv:2411.18641 [hep-ex]} \BibitemShut {NoStop}%
\bibitem [{\citenamefont {De~Romeri}\ \emph {et~al.}(2024{\natexlab{b}})\citenamefont {De~Romeri}, \citenamefont {Papoulias},\ and\ \citenamefont {Ternes}}]{DeRomeri:2024iaw}%
  \BibitemOpen
  \bibfield  {author} {\bibinfo {author} {\bibfnamefont {V.}~\bibnamefont {De~Romeri}}, \bibinfo {author} {\bibfnamefont {D.~K.}\ \bibnamefont {Papoulias}}, \ and\ \bibinfo {author} {\bibfnamefont {C.~A.}\ \bibnamefont {Ternes}},\ }\href@noop {} {\enquote {\bibinfo {title} {{Bounds on new neutrino interactions from the first CE$\nu$NS data at direct detection experiments}},}\ } (\bibinfo {year} {2024}{\natexlab{b}}),\ \Eprint {http://arxiv.org/abs/2411.11749} {arXiv:2411.11749 [hep-ph]} \BibitemShut {NoStop}%
\bibitem [{\citenamefont {Majumdar}\ \emph {et~al.}(2024)\citenamefont {Majumdar}, \citenamefont {Papoulias}, \citenamefont {Prajapati},\ and\ \citenamefont {Srivastava}}]{Majumdar:2024dms}%
  \BibitemOpen
  \bibfield  {author} {\bibinfo {author} {\bibfnamefont {A.}~\bibnamefont {Majumdar}}, \bibinfo {author} {\bibfnamefont {D.~K.}\ \bibnamefont {Papoulias}}, \bibinfo {author} {\bibfnamefont {H.}~\bibnamefont {Prajapati}}, \ and\ \bibinfo {author} {\bibfnamefont {R.}~\bibnamefont {Srivastava}},\ }\href@noop {} {\enquote {\bibinfo {title} {{Constraining low scale Dark Hypercharge symmetry at spallation, reactor and Dark Matter direct detection experiments}},}\ } (\bibinfo {year} {2024}),\ \Eprint {http://arxiv.org/abs/2411.04197} {arXiv:2411.04197 [hep-ph]} \BibitemShut {NoStop}%
\bibitem [{\citenamefont {Pandey}(2024)}]{Pandey:2023arh}%
  \BibitemOpen
  \bibfield  {author} {\bibinfo {author} {\bibfnamefont {V.}~\bibnamefont {Pandey}},\ }\href {\doibase 10.1016/j.ppnp.2023.104078} {\bibfield  {journal} {\bibinfo  {journal} {Prog. Part. Nucl. Phys.}\ }\textbf {\bibinfo {volume} {134}},\ \bibinfo {pages} {104078} (\bibinfo {year} {2024})},\ \Eprint {http://arxiv.org/abs/2309.07840} {arXiv:2309.07840 [hep-ph]} \BibitemShut {NoStop}%
\bibitem [{\citenamefont {Coloma}\ \emph {et~al.}(2023)\citenamefont {Coloma}, \citenamefont {Gonzalez-Garcia}, \citenamefont {Maltoni}, \citenamefont {Pinheiro},\ and\ \citenamefont {Urrea}}]{Coloma:2023ixt}%
  \BibitemOpen
  \bibfield  {author} {\bibinfo {author} {\bibfnamefont {P.}~\bibnamefont {Coloma}}, \bibinfo {author} {\bibfnamefont {M.~C.}\ \bibnamefont {Gonzalez-Garcia}}, \bibinfo {author} {\bibfnamefont {M.}~\bibnamefont {Maltoni}}, \bibinfo {author} {\bibfnamefont {J.~a.~P.}\ \bibnamefont {Pinheiro}}, \ and\ \bibinfo {author} {\bibfnamefont {S.}~\bibnamefont {Urrea}},\ }\href {\doibase 10.1007/JHEP08(2023)032} {\bibfield  {journal} {\bibinfo  {journal} {JHEP}\ }\textbf {\bibinfo {volume} {08}},\ \bibinfo {pages} {032} (\bibinfo {year} {2023})},\ \Eprint {http://arxiv.org/abs/2305.07698} {arXiv:2305.07698 [hep-ph]} \BibitemShut {NoStop}%
\bibitem [{\citenamefont {Aristizabal~Sierra}\ \emph {et~al.}(2024)\citenamefont {Aristizabal~Sierra}, \citenamefont {Mishra},\ and\ \citenamefont {Strigari}}]{AristizabalSierra:2024nwf}%
  \BibitemOpen
  \bibfield  {author} {\bibinfo {author} {\bibfnamefont {D.}~\bibnamefont {Aristizabal~Sierra}}, \bibinfo {author} {\bibfnamefont {N.}~\bibnamefont {Mishra}}, \ and\ \bibinfo {author} {\bibfnamefont {L.}~\bibnamefont {Strigari}},\ }\href@noop {} {\enquote {\bibinfo {title} {{Implications of first neutrino-induced nuclear recoil measurements in direct detection experiments}},}\ } (\bibinfo {year} {2024}),\ \Eprint {http://arxiv.org/abs/2409.02003} {arXiv:2409.02003 [hep-ph]} \BibitemShut {NoStop}%
\bibitem [{\citenamefont {Cadeddu}\ \emph {et~al.}(2018{\natexlab{a}})\citenamefont {Cadeddu}, \citenamefont {Giunti}, \citenamefont {Li},\ and\ \citenamefont {Zhang}}]{Cadeddu:2017etk}%
  \BibitemOpen
  \bibfield  {author} {\bibinfo {author} {\bibfnamefont {M.}~\bibnamefont {Cadeddu}}, \bibinfo {author} {\bibfnamefont {C.}~\bibnamefont {Giunti}}, \bibinfo {author} {\bibfnamefont {Y.~F.}\ \bibnamefont {Li}}, \ and\ \bibinfo {author} {\bibfnamefont {Y.~Y.}\ \bibnamefont {Zhang}},\ }\href {\doibase 10.1103/PhysRevLett.120.072501} {\bibfield  {journal} {\bibinfo  {journal} {Phys. Rev. Lett.}\ }\textbf {\bibinfo {volume} {120}},\ \bibinfo {pages} {072501} (\bibinfo {year} {2018}{\natexlab{a}})},\ \Eprint {http://arxiv.org/abs/1710.02730} {arXiv:1710.02730 [hep-ph]} \BibitemShut {NoStop}%
\bibitem [{\citenamefont {Cadeddu}\ \emph {et~al.}(2018{\natexlab{b}})\citenamefont {Cadeddu}, \citenamefont {Giunti}, \citenamefont {Kouzakov}, \citenamefont {Li}, \citenamefont {Zhang},\ and\ \citenamefont {Studenikin}}]{Cadeddu:2018dux}%
  \BibitemOpen
  \bibfield  {author} {\bibinfo {author} {\bibfnamefont {M.}~\bibnamefont {Cadeddu}}, \bibinfo {author} {\bibfnamefont {C.}~\bibnamefont {Giunti}}, \bibinfo {author} {\bibfnamefont {K.~A.}\ \bibnamefont {Kouzakov}}, \bibinfo {author} {\bibfnamefont {Y.-F.}\ \bibnamefont {Li}}, \bibinfo {author} {\bibfnamefont {Y.-Y.}\ \bibnamefont {Zhang}}, \ and\ \bibinfo {author} {\bibfnamefont {A.~I.}\ \bibnamefont {Studenikin}},\ }\href {\doibase 10.1142/9789811233913_0013} {\bibfield  {journal} {\bibinfo  {journal} {Phys. Rev. D}\ }\textbf {\bibinfo {volume} {98}},\ \bibinfo {pages} {113010} (\bibinfo {year} {2018}{\natexlab{b}})},\ \bibinfo {note} {[Erratum: Phys.Rev.D 101, 059902 (2020)]},\ \Eprint {http://arxiv.org/abs/1810.05606} {arXiv:1810.05606 [hep-ph]} \BibitemShut {NoStop}%
\bibitem [{\citenamefont {Cadeddu}\ \emph {et~al.}(2020{\natexlab{a}})\citenamefont {Cadeddu}, \citenamefont {Dordei}, \citenamefont {Giunti}, \citenamefont {Li},\ and\ \citenamefont {Zhang}}]{Cadeddu:2019eta}%
  \BibitemOpen
  \bibfield  {author} {\bibinfo {author} {\bibfnamefont {M.}~\bibnamefont {Cadeddu}}, \bibinfo {author} {\bibfnamefont {F.}~\bibnamefont {Dordei}}, \bibinfo {author} {\bibfnamefont {C.}~\bibnamefont {Giunti}}, \bibinfo {author} {\bibfnamefont {Y.~F.}\ \bibnamefont {Li}}, \ and\ \bibinfo {author} {\bibfnamefont {Y.~Y.}\ \bibnamefont {Zhang}},\ }\href {\doibase 10.1103/PhysRevD.101.033004} {\bibfield  {journal} {\bibinfo  {journal} {Phys. Rev. D}\ }\textbf {\bibinfo {volume} {101}},\ \bibinfo {pages} {033004} (\bibinfo {year} {2020}{\natexlab{a}})},\ \Eprint {http://arxiv.org/abs/1908.06045} {arXiv:1908.06045 [hep-ph]} \BibitemShut {NoStop}%
\bibitem [{\citenamefont {Cadeddu}\ \emph {et~al.}(2020{\natexlab{b}})\citenamefont {Cadeddu}, \citenamefont {Dordei}, \citenamefont {Giunti}, \citenamefont {Li}, \citenamefont {Picciau},\ and\ \citenamefont {Zhang}}]{Cadeddu:2020lky}%
  \BibitemOpen
  \bibfield  {author} {\bibinfo {author} {\bibfnamefont {M.}~\bibnamefont {Cadeddu}}, \bibinfo {author} {\bibfnamefont {F.}~\bibnamefont {Dordei}}, \bibinfo {author} {\bibfnamefont {C.}~\bibnamefont {Giunti}}, \bibinfo {author} {\bibfnamefont {Y.~F.}\ \bibnamefont {Li}}, \bibinfo {author} {\bibfnamefont {E.}~\bibnamefont {Picciau}}, \ and\ \bibinfo {author} {\bibfnamefont {Y.~Y.}\ \bibnamefont {Zhang}},\ }\href {\doibase 10.1103/PhysRevD.102.015030} {\bibfield  {journal} {\bibinfo  {journal} {Phys. Rev. D}\ }\textbf {\bibinfo {volume} {102}},\ \bibinfo {pages} {015030} (\bibinfo {year} {2020}{\natexlab{b}})},\ \Eprint {http://arxiv.org/abs/2005.01645} {arXiv:2005.01645 [hep-ph]} \BibitemShut {NoStop}%
\bibitem [{\citenamefont {Cadeddu}\ and\ \citenamefont {Dordei}(2019)}]{Cadeddu:2018izq}%
  \BibitemOpen
  \bibfield  {author} {\bibinfo {author} {\bibfnamefont {M.}~\bibnamefont {Cadeddu}}\ and\ \bibinfo {author} {\bibfnamefont {F.}~\bibnamefont {Dordei}},\ }\href {\doibase 10.1103/PhysRevD.99.033010} {\bibfield  {journal} {\bibinfo  {journal} {Phys. Rev. D}\ }\textbf {\bibinfo {volume} {99}},\ \bibinfo {pages} {033010} (\bibinfo {year} {2019})},\ \Eprint {http://arxiv.org/abs/1808.10202} {arXiv:1808.10202 [hep-ph]} \BibitemShut {NoStop}%
\bibitem [{\citenamefont {Cadeddu}\ \emph {et~al.}(2021{\natexlab{a}})\citenamefont {Cadeddu}, \citenamefont {Cargioli}, \citenamefont {Dordei}, \citenamefont {Giunti}, \citenamefont {Li}, \citenamefont {Picciau},\ and\ \citenamefont {Zhang}}]{Cadeddu:2020nbr}%
  \BibitemOpen
  \bibfield  {author} {\bibinfo {author} {\bibfnamefont {M.}~\bibnamefont {Cadeddu}}, \bibinfo {author} {\bibfnamefont {N.}~\bibnamefont {Cargioli}}, \bibinfo {author} {\bibfnamefont {F.}~\bibnamefont {Dordei}}, \bibinfo {author} {\bibfnamefont {C.}~\bibnamefont {Giunti}}, \bibinfo {author} {\bibfnamefont {Y.~F.}\ \bibnamefont {Li}}, \bibinfo {author} {\bibfnamefont {E.}~\bibnamefont {Picciau}}, \ and\ \bibinfo {author} {\bibfnamefont {Y.~Y.}\ \bibnamefont {Zhang}},\ }\href {\doibase 10.1007/JHEP01(2021)116} {\bibfield  {journal} {\bibinfo  {journal} {JHEP}\ }\textbf {\bibinfo {volume} {01}},\ \bibinfo {pages} {116} (\bibinfo {year} {2021}{\natexlab{a}})},\ \Eprint {http://arxiv.org/abs/2008.05022} {arXiv:2008.05022 [hep-ph]} \BibitemShut {NoStop}%
\bibitem [{\citenamefont {Cadeddu}\ \emph {et~al.}(2021{\natexlab{b}})\citenamefont {Cadeddu}, \citenamefont {Cargioli}, \citenamefont {Dordei}, \citenamefont {Giunti}, \citenamefont {Li}, \citenamefont {Picciau}, \citenamefont {Ternes},\ and\ \citenamefont {Zhang}}]{Cadeddu:2021ijh}%
  \BibitemOpen
  \bibfield  {author} {\bibinfo {author} {\bibfnamefont {M.}~\bibnamefont {Cadeddu}}, \bibinfo {author} {\bibfnamefont {N.}~\bibnamefont {Cargioli}}, \bibinfo {author} {\bibfnamefont {F.}~\bibnamefont {Dordei}}, \bibinfo {author} {\bibfnamefont {C.}~\bibnamefont {Giunti}}, \bibinfo {author} {\bibfnamefont {Y.~F.}\ \bibnamefont {Li}}, \bibinfo {author} {\bibfnamefont {E.}~\bibnamefont {Picciau}}, \bibinfo {author} {\bibfnamefont {C.~A.}\ \bibnamefont {Ternes}}, \ and\ \bibinfo {author} {\bibfnamefont {Y.~Y.}\ \bibnamefont {Zhang}},\ }\href {\doibase 10.1103/PhysRevC.104.065502} {\bibfield  {journal} {\bibinfo  {journal} {Phys. Rev. C}\ }\textbf {\bibinfo {volume} {104}},\ \bibinfo {pages} {065502} (\bibinfo {year} {2021}{\natexlab{b}})},\ \Eprint {http://arxiv.org/abs/2102.06153} {arXiv:2102.06153 [hep-ph]} \BibitemShut {NoStop}%
\bibitem [{\citenamefont {Atzori~Corona}\ \emph {et~al.}(2022{\natexlab{a}})\citenamefont {Atzori~Corona}, \citenamefont {Cadeddu}, \citenamefont {Cargioli}, \citenamefont {Dordei}, \citenamefont {Giunti}, \citenamefont {Li}, \citenamefont {Picciau}, \citenamefont {Ternes},\ and\ \citenamefont {Zhang}}]{AtzoriCorona:2022moj}%
  \BibitemOpen
  \bibfield  {author} {\bibinfo {author} {\bibfnamefont {M.}~\bibnamefont {Atzori~Corona}}, \bibinfo {author} {\bibfnamefont {M.}~\bibnamefont {Cadeddu}}, \bibinfo {author} {\bibfnamefont {N.}~\bibnamefont {Cargioli}}, \bibinfo {author} {\bibfnamefont {F.}~\bibnamefont {Dordei}}, \bibinfo {author} {\bibfnamefont {C.}~\bibnamefont {Giunti}}, \bibinfo {author} {\bibfnamefont {Y.~F.}\ \bibnamefont {Li}}, \bibinfo {author} {\bibfnamefont {E.}~\bibnamefont {Picciau}}, \bibinfo {author} {\bibfnamefont {C.~A.}\ \bibnamefont {Ternes}}, \ and\ \bibinfo {author} {\bibfnamefont {Y.~Y.}\ \bibnamefont {Zhang}},\ }\href {\doibase 10.1007/JHEP05(2022)109} {\bibfield  {journal} {\bibinfo  {journal} {JHEP}\ }\textbf {\bibinfo {volume} {05}},\ \bibinfo {pages} {109} (\bibinfo {year} {2022}{\natexlab{a}})},\ \Eprint {http://arxiv.org/abs/2202.11002} {arXiv:2202.11002 [hep-ph]} \BibitemShut {NoStop}%
\bibitem [{\citenamefont {Atzori~Corona}\ \emph {et~al.}(2022{\natexlab{b}})\citenamefont {Atzori~Corona}, \citenamefont {Cadeddu}, \citenamefont {Cargioli}, \citenamefont {Dordei}, \citenamefont {Giunti}, \citenamefont {Li}, \citenamefont {Ternes},\ and\ \citenamefont {Zhang}}]{AtzoriCorona:2022qrf}%
  \BibitemOpen
  \bibfield  {author} {\bibinfo {author} {\bibfnamefont {M.}~\bibnamefont {Atzori~Corona}}, \bibinfo {author} {\bibfnamefont {M.}~\bibnamefont {Cadeddu}}, \bibinfo {author} {\bibfnamefont {N.}~\bibnamefont {Cargioli}}, \bibinfo {author} {\bibfnamefont {F.}~\bibnamefont {Dordei}}, \bibinfo {author} {\bibfnamefont {C.}~\bibnamefont {Giunti}}, \bibinfo {author} {\bibfnamefont {Y.~F.}\ \bibnamefont {Li}}, \bibinfo {author} {\bibfnamefont {C.~A.}\ \bibnamefont {Ternes}}, \ and\ \bibinfo {author} {\bibfnamefont {Y.~Y.}\ \bibnamefont {Zhang}},\ }\href {\doibase 10.1007/JHEP09(2022)164} {\bibfield  {journal} {\bibinfo  {journal} {JHEP}\ }\textbf {\bibinfo {volume} {09}},\ \bibinfo {pages} {164} (\bibinfo {year} {2022}{\natexlab{b}})},\ \Eprint {http://arxiv.org/abs/2205.09484} {arXiv:2205.09484 [hep-ph]} \BibitemShut {NoStop}%
\bibitem [{\citenamefont {Atzori~Corona}\ \emph {et~al.}(2023)\citenamefont {Atzori~Corona}, \citenamefont {Cadeddu}, \citenamefont {Cargioli}, \citenamefont {Dordei}, \citenamefont {Giunti},\ and\ \citenamefont {Masia}}]{AtzoriCorona:2023ktl}%
  \BibitemOpen
  \bibfield  {author} {\bibinfo {author} {\bibfnamefont {M.}~\bibnamefont {Atzori~Corona}}, \bibinfo {author} {\bibfnamefont {M.}~\bibnamefont {Cadeddu}}, \bibinfo {author} {\bibfnamefont {N.}~\bibnamefont {Cargioli}}, \bibinfo {author} {\bibfnamefont {F.}~\bibnamefont {Dordei}}, \bibinfo {author} {\bibfnamefont {C.}~\bibnamefont {Giunti}}, \ and\ \bibinfo {author} {\bibfnamefont {G.}~\bibnamefont {Masia}},\ }\href {\doibase 10.1140/epjc/s10052-023-11849-5} {\bibfield  {journal} {\bibinfo  {journal} {Eur. Phys. J. C}\ }\textbf {\bibinfo {volume} {83}},\ \bibinfo {pages} {683} (\bibinfo {year} {2023})},\ \Eprint {http://arxiv.org/abs/2303.09360} {arXiv:2303.09360 [nucl-ex]} \BibitemShut {NoStop}%
\bibitem [{\citenamefont {Atzori~Corona}\ \emph {et~al.}(2024{\natexlab{a}})\citenamefont {Atzori~Corona}, \citenamefont {Cadeddu}, \citenamefont {Cargioli}, \citenamefont {Dordei},\ and\ \citenamefont {Giunti}}]{AtzoriCorona:2024rtv}%
  \BibitemOpen
  \bibfield  {author} {\bibinfo {author} {\bibfnamefont {M.}~\bibnamefont {Atzori~Corona}}, \bibinfo {author} {\bibfnamefont {M.}~\bibnamefont {Cadeddu}}, \bibinfo {author} {\bibfnamefont {N.}~\bibnamefont {Cargioli}}, \bibinfo {author} {\bibfnamefont {F.}~\bibnamefont {Dordei}}, \ and\ \bibinfo {author} {\bibfnamefont {C.}~\bibnamefont {Giunti}},\ }\href {\doibase 10.1007/JHEP05(2024)271} {\bibfield  {journal} {\bibinfo  {journal} {JHEP}\ }\textbf {\bibinfo {volume} {05}},\ \bibinfo {pages} {271} (\bibinfo {year} {2024}{\natexlab{a}})},\ \Eprint {http://arxiv.org/abs/2402.16709} {arXiv:2402.16709 [hep-ph]} \BibitemShut {NoStop}%
\bibitem [{\citenamefont {Coloma}\ \emph {et~al.}(2017)\citenamefont {Coloma}, \citenamefont {Gonzalez-Garcia}, \citenamefont {Maltoni},\ and\ \citenamefont {Schwetz}}]{Coloma:2017ncl}%
  \BibitemOpen
  \bibfield  {author} {\bibinfo {author} {\bibfnamefont {P.}~\bibnamefont {Coloma}}, \bibinfo {author} {\bibfnamefont {M.~C.}\ \bibnamefont {Gonzalez-Garcia}}, \bibinfo {author} {\bibfnamefont {M.}~\bibnamefont {Maltoni}}, \ and\ \bibinfo {author} {\bibfnamefont {T.}~\bibnamefont {Schwetz}},\ }\href {\doibase 10.1103/PhysRevD.96.115007} {\bibfield  {journal} {\bibinfo  {journal} {Phys. Rev. D}\ }\textbf {\bibinfo {volume} {96}},\ \bibinfo {pages} {115007} (\bibinfo {year} {2017})},\ \Eprint {http://arxiv.org/abs/1708.02899} {arXiv:1708.02899 [hep-ph]} \BibitemShut {NoStop}%
\bibitem [{\citenamefont {Liao}\ and\ \citenamefont {Marfatia}(2017)}]{Liao:2017uzy}%
  \BibitemOpen
  \bibfield  {author} {\bibinfo {author} {\bibfnamefont {J.}~\bibnamefont {Liao}}\ and\ \bibinfo {author} {\bibfnamefont {D.}~\bibnamefont {Marfatia}},\ }\href {\doibase 10.1016/j.physletb.2017.10.046} {\bibfield  {journal} {\bibinfo  {journal} {Phys. Lett. B}\ }\textbf {\bibinfo {volume} {775}},\ \bibinfo {pages} {54} (\bibinfo {year} {2017})},\ \Eprint {http://arxiv.org/abs/1708.04255} {arXiv:1708.04255 [hep-ph]} \BibitemShut {NoStop}%
\bibitem [{\citenamefont {Lindner}\ \emph {et~al.}(2017)\citenamefont {Lindner}, \citenamefont {Rodejohann},\ and\ \citenamefont {Xu}}]{Lindner:2016wff}%
  \BibitemOpen
  \bibfield  {author} {\bibinfo {author} {\bibfnamefont {M.}~\bibnamefont {Lindner}}, \bibinfo {author} {\bibfnamefont {W.}~\bibnamefont {Rodejohann}}, \ and\ \bibinfo {author} {\bibfnamefont {X.-J.}\ \bibnamefont {Xu}},\ }\href {\doibase 10.1007/JHEP03(2017)097} {\bibfield  {journal} {\bibinfo  {journal} {JHEP}\ }\textbf {\bibinfo {volume} {03}},\ \bibinfo {pages} {097} (\bibinfo {year} {2017})},\ \Eprint {http://arxiv.org/abs/1612.04150} {arXiv:1612.04150 [hep-ph]} \BibitemShut {NoStop}%
\bibitem [{\citenamefont {Giunti}(2020)}]{Giunti:2019xpr}%
  \BibitemOpen
  \bibfield  {author} {\bibinfo {author} {\bibfnamefont {C.}~\bibnamefont {Giunti}},\ }\href {\doibase 10.1103/PhysRevD.101.035039} {\bibfield  {journal} {\bibinfo  {journal} {Phys. Rev. D}\ }\textbf {\bibinfo {volume} {101}},\ \bibinfo {pages} {035039} (\bibinfo {year} {2020})},\ \Eprint {http://arxiv.org/abs/1909.00466} {arXiv:1909.00466 [hep-ph]} \BibitemShut {NoStop}%
\bibitem [{\citenamefont {Denton}\ \emph {et~al.}(2018)\citenamefont {Denton}, \citenamefont {Farzan},\ and\ \citenamefont {Shoemaker}}]{Denton:2018xmq}%
  \BibitemOpen
  \bibfield  {author} {\bibinfo {author} {\bibfnamefont {P.~B.}\ \bibnamefont {Denton}}, \bibinfo {author} {\bibfnamefont {Y.}~\bibnamefont {Farzan}}, \ and\ \bibinfo {author} {\bibfnamefont {I.~M.}\ \bibnamefont {Shoemaker}},\ }\href {\doibase 10.1007/JHEP07(2018)037} {\bibfield  {journal} {\bibinfo  {journal} {JHEP}\ }\textbf {\bibinfo {volume} {07}},\ \bibinfo {pages} {037} (\bibinfo {year} {2018})},\ \Eprint {http://arxiv.org/abs/1804.03660} {arXiv:1804.03660 [hep-ph]} \BibitemShut {NoStop}%
\bibitem [{\citenamefont {Aristizabal~Sierra}\ \emph {et~al.}(2018)\citenamefont {Aristizabal~Sierra}, \citenamefont {De~Romeri},\ and\ \citenamefont {Rojas}}]{AristizabalSierra:2018eqm}%
  \BibitemOpen
  \bibfield  {author} {\bibinfo {author} {\bibfnamefont {D.}~\bibnamefont {Aristizabal~Sierra}}, \bibinfo {author} {\bibfnamefont {V.}~\bibnamefont {De~Romeri}}, \ and\ \bibinfo {author} {\bibfnamefont {N.}~\bibnamefont {Rojas}},\ }\href {\doibase 10.1103/PhysRevD.98.075018} {\bibfield  {journal} {\bibinfo  {journal} {Phys. Rev. D}\ }\textbf {\bibinfo {volume} {98}},\ \bibinfo {pages} {075018} (\bibinfo {year} {2018})},\ \Eprint {http://arxiv.org/abs/1806.07424} {arXiv:1806.07424 [hep-ph]} \BibitemShut {NoStop}%
\bibitem [{\citenamefont {Miranda}\ \emph {et~al.}(2020)\citenamefont {Miranda}, \citenamefont {Papoulias}, \citenamefont {Sanchez~Garcia}, \citenamefont {Sanders}, \citenamefont {T\'ortola},\ and\ \citenamefont {Valle}}]{Miranda:2020tif}%
  \BibitemOpen
  \bibfield  {author} {\bibinfo {author} {\bibfnamefont {O.~G.}\ \bibnamefont {Miranda}}, \bibinfo {author} {\bibfnamefont {D.~K.}\ \bibnamefont {Papoulias}}, \bibinfo {author} {\bibfnamefont {G.}~\bibnamefont {Sanchez~Garcia}}, \bibinfo {author} {\bibfnamefont {O.}~\bibnamefont {Sanders}}, \bibinfo {author} {\bibfnamefont {M.}~\bibnamefont {T\'ortola}}, \ and\ \bibinfo {author} {\bibfnamefont {J.~W.~F.}\ \bibnamefont {Valle}},\ }\href {\doibase 10.1007/JHEP05(2020)130} {\bibfield  {journal} {\bibinfo  {journal} {JHEP}\ }\textbf {\bibinfo {volume} {05}},\ \bibinfo {pages} {130} (\bibinfo {year} {2020})},\ \bibinfo {note} {[Erratum: JHEP 01, 067 (2021)]},\ \Eprint {http://arxiv.org/abs/2003.12050} {arXiv:2003.12050 [hep-ph]} \BibitemShut {NoStop}%
\bibitem [{\citenamefont {Banerjee}\ \emph {et~al.}(2021)\citenamefont {Banerjee}, \citenamefont {Dutta},\ and\ \citenamefont {Roy}}]{Banerjee:2021laz}%
  \BibitemOpen
  \bibfield  {author} {\bibinfo {author} {\bibfnamefont {H.}~\bibnamefont {Banerjee}}, \bibinfo {author} {\bibfnamefont {B.}~\bibnamefont {Dutta}}, \ and\ \bibinfo {author} {\bibfnamefont {S.}~\bibnamefont {Roy}},\ }\href {\doibase 10.1103/PhysRevD.104.015015} {\bibfield  {journal} {\bibinfo  {journal} {Phys. Rev. D}\ }\textbf {\bibinfo {volume} {104}},\ \bibinfo {pages} {015015} (\bibinfo {year} {2021})},\ \Eprint {http://arxiv.org/abs/2103.10196} {arXiv:2103.10196 [hep-ph]} \BibitemShut {NoStop}%
\bibitem [{\citenamefont {Papoulias}\ \emph {et~al.}(2020)\citenamefont {Papoulias}, \citenamefont {Kosmas}, \citenamefont {Sahu}, \citenamefont {Kota},\ and\ \citenamefont {Hota}}]{Papoulias:2019lfi}%
  \BibitemOpen
  \bibfield  {author} {\bibinfo {author} {\bibfnamefont {D.~K.}\ \bibnamefont {Papoulias}}, \bibinfo {author} {\bibfnamefont {T.~S.}\ \bibnamefont {Kosmas}}, \bibinfo {author} {\bibfnamefont {R.}~\bibnamefont {Sahu}}, \bibinfo {author} {\bibfnamefont {V.~K.~B.}\ \bibnamefont {Kota}}, \ and\ \bibinfo {author} {\bibfnamefont {M.}~\bibnamefont {Hota}},\ }\href {\doibase 10.1016/j.physletb.2019.135133} {\bibfield  {journal} {\bibinfo  {journal} {Phys. Lett. B}\ }\textbf {\bibinfo {volume} {800}},\ \bibinfo {pages} {135133} (\bibinfo {year} {2020})},\ \Eprint {http://arxiv.org/abs/1903.03722} {arXiv:1903.03722 [hep-ph]} \BibitemShut {NoStop}%
\bibitem [{\citenamefont {Papoulias}\ and\ \citenamefont {Kosmas}(2018)}]{Papoulias:2017qdn}%
  \BibitemOpen
  \bibfield  {author} {\bibinfo {author} {\bibfnamefont {D.~K.}\ \bibnamefont {Papoulias}}\ and\ \bibinfo {author} {\bibfnamefont {T.~S.}\ \bibnamefont {Kosmas}},\ }\href {\doibase 10.1103/PhysRevD.97.033003} {\bibfield  {journal} {\bibinfo  {journal} {Phys. Rev. D}\ }\textbf {\bibinfo {volume} {97}},\ \bibinfo {pages} {033003} (\bibinfo {year} {2018})},\ \Eprint {http://arxiv.org/abs/1711.09773} {arXiv:1711.09773 [hep-ph]} \BibitemShut {NoStop}%
\bibitem [{\citenamefont {Dutta}\ \emph {et~al.}(2020)\citenamefont {Dutta}, \citenamefont {Kim}, \citenamefont {Liao}, \citenamefont {Park}, \citenamefont {Shin},\ and\ \citenamefont {Strigari}}]{Dutta:2019nbn}%
  \BibitemOpen
  \bibfield  {author} {\bibinfo {author} {\bibfnamefont {B.}~\bibnamefont {Dutta}}, \bibinfo {author} {\bibfnamefont {D.}~\bibnamefont {Kim}}, \bibinfo {author} {\bibfnamefont {S.}~\bibnamefont {Liao}}, \bibinfo {author} {\bibfnamefont {J.-C.}\ \bibnamefont {Park}}, \bibinfo {author} {\bibfnamefont {S.}~\bibnamefont {Shin}}, \ and\ \bibinfo {author} {\bibfnamefont {L.~E.}\ \bibnamefont {Strigari}},\ }\href {\doibase 10.1103/PhysRevLett.124.121802} {\bibfield  {journal} {\bibinfo  {journal} {Phys. Rev. Lett.}\ }\textbf {\bibinfo {volume} {124}},\ \bibinfo {pages} {121802} (\bibinfo {year} {2020})},\ \Eprint {http://arxiv.org/abs/1906.10745} {arXiv:1906.10745 [hep-ph]} \BibitemShut {NoStop}%
\bibitem [{\citenamefont {Abdullah}\ \emph {et~al.}(2018)\citenamefont {Abdullah}, \citenamefont {Dent}, \citenamefont {Dutta}, \citenamefont {Kane}, \citenamefont {Liao},\ and\ \citenamefont {Strigari}}]{Abdullah:2018ykz}%
  \BibitemOpen
  \bibfield  {author} {\bibinfo {author} {\bibfnamefont {M.}~\bibnamefont {Abdullah}}, \bibinfo {author} {\bibfnamefont {J.~B.}\ \bibnamefont {Dent}}, \bibinfo {author} {\bibfnamefont {B.}~\bibnamefont {Dutta}}, \bibinfo {author} {\bibfnamefont {G.~L.}\ \bibnamefont {Kane}}, \bibinfo {author} {\bibfnamefont {S.}~\bibnamefont {Liao}}, \ and\ \bibinfo {author} {\bibfnamefont {L.~E.}\ \bibnamefont {Strigari}},\ }\href {\doibase 10.1103/PhysRevD.98.015005} {\bibfield  {journal} {\bibinfo  {journal} {Phys. Rev. D}\ }\textbf {\bibinfo {volume} {98}},\ \bibinfo {pages} {015005} (\bibinfo {year} {2018})},\ \Eprint {http://arxiv.org/abs/1803.01224} {arXiv:1803.01224 [hep-ph]} \BibitemShut {NoStop}%
\bibitem [{\citenamefont {Ge}\ and\ \citenamefont {Shoemaker}(2018)}]{Ge:2017mcq}%
  \BibitemOpen
  \bibfield  {author} {\bibinfo {author} {\bibfnamefont {S.-F.}\ \bibnamefont {Ge}}\ and\ \bibinfo {author} {\bibfnamefont {I.~M.}\ \bibnamefont {Shoemaker}},\ }\href {\doibase 10.1007/JHEP11(2018)066} {\bibfield  {journal} {\bibinfo  {journal} {JHEP}\ }\textbf {\bibinfo {volume} {11}},\ \bibinfo {pages} {066} (\bibinfo {year} {2018})},\ \Eprint {http://arxiv.org/abs/1710.10889} {arXiv:1710.10889 [hep-ph]} \BibitemShut {NoStop}%
\bibitem [{\citenamefont {Miranda}\ \emph {et~al.}(2021)\citenamefont {Miranda}, \citenamefont {Papoulias}, \citenamefont {Sanders}, \citenamefont {T\'ortola},\ and\ \citenamefont {Valle}}]{Miranda:2021kre}%
  \BibitemOpen
  \bibfield  {author} {\bibinfo {author} {\bibfnamefont {O.~G.}\ \bibnamefont {Miranda}}, \bibinfo {author} {\bibfnamefont {D.~K.}\ \bibnamefont {Papoulias}}, \bibinfo {author} {\bibfnamefont {O.}~\bibnamefont {Sanders}}, \bibinfo {author} {\bibfnamefont {M.}~\bibnamefont {T\'ortola}}, \ and\ \bibinfo {author} {\bibfnamefont {J.~W.~F.}\ \bibnamefont {Valle}},\ }\href {\doibase 10.1007/JHEP12(2021)191} {\bibfield  {journal} {\bibinfo  {journal} {JHEP}\ }\textbf {\bibinfo {volume} {12}},\ \bibinfo {pages} {191} (\bibinfo {year} {2021})},\ \Eprint {http://arxiv.org/abs/2109.09545} {arXiv:2109.09545 [hep-ph]} \BibitemShut {NoStop}%
\bibitem [{\citenamefont {Flores}\ \emph {et~al.}(2020)\citenamefont {Flores}, \citenamefont {Nath},\ and\ \citenamefont {Peinado}}]{Flores:2020lji}%
  \BibitemOpen
  \bibfield  {author} {\bibinfo {author} {\bibfnamefont {L.~J.}\ \bibnamefont {Flores}}, \bibinfo {author} {\bibfnamefont {N.}~\bibnamefont {Nath}}, \ and\ \bibinfo {author} {\bibfnamefont {E.}~\bibnamefont {Peinado}},\ }\href {\doibase 10.1007/JHEP06(2020)045} {\bibfield  {journal} {\bibinfo  {journal} {JHEP}\ }\textbf {\bibinfo {volume} {06}},\ \bibinfo {pages} {045} (\bibinfo {year} {2020})},\ \Eprint {http://arxiv.org/abs/2002.12342} {arXiv:2002.12342 [hep-ph]} \BibitemShut {NoStop}%
\bibitem [{\citenamefont {Farzan}\ \emph {et~al.}(2018)\citenamefont {Farzan}, \citenamefont {Lindner}, \citenamefont {Rodejohann},\ and\ \citenamefont {Xu}}]{Farzan:2018gtr}%
  \BibitemOpen
  \bibfield  {author} {\bibinfo {author} {\bibfnamefont {Y.}~\bibnamefont {Farzan}}, \bibinfo {author} {\bibfnamefont {M.}~\bibnamefont {Lindner}}, \bibinfo {author} {\bibfnamefont {W.}~\bibnamefont {Rodejohann}}, \ and\ \bibinfo {author} {\bibfnamefont {X.-J.}\ \bibnamefont {Xu}},\ }\href {\doibase 10.1007/JHEP05(2018)066} {\bibfield  {journal} {\bibinfo  {journal} {JHEP}\ }\textbf {\bibinfo {volume} {05}},\ \bibinfo {pages} {066} (\bibinfo {year} {2018})},\ \Eprint {http://arxiv.org/abs/1802.05171} {arXiv:1802.05171 [hep-ph]} \BibitemShut {NoStop}%
\bibitem [{\citenamefont {Brdar}\ \emph {et~al.}(2018)\citenamefont {Brdar}, \citenamefont {Rodejohann},\ and\ \citenamefont {Xu}}]{Brdar:2018qqj}%
  \BibitemOpen
  \bibfield  {author} {\bibinfo {author} {\bibfnamefont {V.}~\bibnamefont {Brdar}}, \bibinfo {author} {\bibfnamefont {W.}~\bibnamefont {Rodejohann}}, \ and\ \bibinfo {author} {\bibfnamefont {X.-J.}\ \bibnamefont {Xu}},\ }\href {\doibase 10.1007/JHEP12(2018)024} {\bibfield  {journal} {\bibinfo  {journal} {JHEP}\ }\textbf {\bibinfo {volume} {12}},\ \bibinfo {pages} {024} (\bibinfo {year} {2018})},\ \Eprint {http://arxiv.org/abs/1810.03626} {arXiv:1810.03626 [hep-ph]} \BibitemShut {NoStop}%
\bibitem [{\citenamefont {Adamski}\ \emph {et~al.}(2025)\citenamefont {Adamski} \emph {et~al.}}]{COHERENT:2025vuz}%
  \BibitemOpen
  \bibfield  {author} {\bibinfo {author} {\bibfnamefont {S.}~\bibnamefont {Adamski}} \emph {et~al.} (\bibinfo {collaboration} {COHERENT}),\ }\href {\doibase 10.1103/PhysRevLett.134.231801} {\bibfield  {journal} {\bibinfo  {journal} {Phys. Rev. Lett.}\ }\textbf {\bibinfo {volume} {134}},\ \bibinfo {pages} {231801} (\bibinfo {year} {2025})}\BibitemShut {NoStop}%
\bibitem [{\citenamefont {Ackermann}\ \emph {et~al.}(2025)\citenamefont {Ackermann} \emph {et~al.}}]{Ackermann:2025obx}%
  \BibitemOpen
  \bibfield  {author} {\bibinfo {author} {\bibfnamefont {N.}~\bibnamefont {Ackermann}} \emph {et~al.},\ }\href@noop {} {\enquote {\bibinfo {title} {{First observation of reactor antineutrinos by coherent scattering}},}\ } (\bibinfo {year} {2025}),\ \Eprint {http://arxiv.org/abs/2501.05206} {arXiv:2501.05206 [hep-ex]} \BibitemShut {NoStop}%
\bibitem [{\citenamefont {Karmakar}\ \emph {et~al.}(2025)\citenamefont {Karmakar} \emph {et~al.}}]{TEXONO:2024vfk}%
  \BibitemOpen
  \bibfield  {author} {\bibinfo {author} {\bibfnamefont {S.}~\bibnamefont {Karmakar}} \emph {et~al.} (\bibinfo {collaboration} {TEXONO}),\ }\href {\doibase 10.1103/PhysRevLett.134.121802} {\bibfield  {journal} {\bibinfo  {journal} {Phys. Rev. Lett.}\ }\textbf {\bibinfo {volume} {134}},\ \bibinfo {pages} {121802} (\bibinfo {year} {2025})},\ \Eprint {http://arxiv.org/abs/2411.18812} {arXiv:2411.18812 [nucl-ex]} \BibitemShut {NoStop}%
\bibitem [{\citenamefont {Belov}\ \emph {et~al.}(2025)\citenamefont {Belov} \emph {et~al.}}]{nuGeN:2025mla}%
  \BibitemOpen
  \bibfield  {author} {\bibinfo {author} {\bibfnamefont {V.}~\bibnamefont {Belov}} \emph {et~al.} (\bibinfo {collaboration} {nuGeN}),\ }\href {\doibase 10.1088/1674-1137/adb9c8} {\  (\bibinfo {year} {2025}),\ 10.1088/1674-1137/adb9c8},\ \Eprint {http://arxiv.org/abs/2502.18502} {arXiv:2502.18502 [hep-ex]} \BibitemShut {NoStop}%
\bibitem [{\citenamefont {Berglund}\ and\ \citenamefont {Wieser}(2011)}]{BerglundWieser+2011+397+410}%
  \BibitemOpen
  \bibfield  {author} {\bibinfo {author} {\bibfnamefont {M.}~\bibnamefont {Berglund}}\ and\ \bibinfo {author} {\bibfnamefont {M.~E.}\ \bibnamefont {Wieser}},\ }\href {\doibase doi:10.1351/PAC-REP-10-06-02} {\bibfield  {journal} {\bibinfo  {journal} {Pure and Applied Chemistry}\ }\textbf {\bibinfo {volume} {83}},\ \bibinfo {pages} {397} (\bibinfo {year} {2011})}\BibitemShut {NoStop}%
\bibitem [{\citenamefont {Erler}\ and\ \citenamefont {Su}(2013)}]{Erler:2013xha}%
  \BibitemOpen
  \bibfield  {author} {\bibinfo {author} {\bibfnamefont {J.}~\bibnamefont {Erler}}\ and\ \bibinfo {author} {\bibfnamefont {S.}~\bibnamefont {Su}},\ }\href {\doibase 10.1016/j.ppnp.2013.03.004} {\bibfield  {journal} {\bibinfo  {journal} {Prog. Part. Nucl. Phys.}\ }\textbf {\bibinfo {volume} {71}},\ \bibinfo {pages} {119} (\bibinfo {year} {2013})},\ \Eprint {http://arxiv.org/abs/1303.5522} {arXiv:1303.5522 [hep-ph]} \BibitemShut {NoStop}%
\bibitem [{\citenamefont {Navas}(2024)}]{PhysRevD.110.030001}%
  \BibitemOpen
  \bibfield  {author} {\bibinfo {author} {\bibfnamefont {S.~e.~a.}\ \bibnamefont {Navas}} (\bibinfo {collaboration} {Particle Data Group Collaboration}),\ }\href {\doibase 10.1103/PhysRevD.110.030001} {\bibfield  {journal} {\bibinfo  {journal} {Phys. Rev. D}\ }\textbf {\bibinfo {volume} {110}},\ \bibinfo {pages} {030001} (\bibinfo {year} {2024})}\BibitemShut {NoStop}%
\bibitem [{\citenamefont {Giunti}\ \emph {et~al.}(2024)\citenamefont {Giunti}, \citenamefont {Kouzakov}, \citenamefont {Li},\ and\ \citenamefont {Studenikin}}]{Giunti:2024gec}%
  \BibitemOpen
  \bibfield  {author} {\bibinfo {author} {\bibfnamefont {C.}~\bibnamefont {Giunti}}, \bibinfo {author} {\bibfnamefont {K.}~\bibnamefont {Kouzakov}}, \bibinfo {author} {\bibfnamefont {Y.-F.}\ \bibnamefont {Li}}, \ and\ \bibinfo {author} {\bibfnamefont {A.}~\bibnamefont {Studenikin}},\ }\href {\doibase 10.1146/annurev-nucl-102122-023242} {\enquote {\bibinfo {title} {{Neutrino Electromagnetic Properties}},}\ } (\bibinfo {year} {2024}),\ \Eprint {http://arxiv.org/abs/2411.03122} {arXiv:2411.03122 [hep-ph]} \BibitemShut {NoStop}%
\bibitem [{\citenamefont {Bernabeu}\ \emph {et~al.}(2000)\citenamefont {Bernabeu}, \citenamefont {Cabral-Rosetti}, \citenamefont {Papavassiliou},\ and\ \citenamefont {Vidal}}]{Bernabeu:2000hf}%
  \BibitemOpen
  \bibfield  {author} {\bibinfo {author} {\bibfnamefont {J.}~\bibnamefont {Bernabeu}}, \bibinfo {author} {\bibfnamefont {L.~G.}\ \bibnamefont {Cabral-Rosetti}}, \bibinfo {author} {\bibfnamefont {J.}~\bibnamefont {Papavassiliou}}, \ and\ \bibinfo {author} {\bibfnamefont {J.}~\bibnamefont {Vidal}},\ }\href {\doibase 10.1103/PhysRevD.62.113012} {\bibfield  {journal} {\bibinfo  {journal} {Phys. Rev. D}\ }\textbf {\bibinfo {volume} {62}},\ \bibinfo {pages} {113012} (\bibinfo {year} {2000})},\ \Eprint {http://arxiv.org/abs/hep-ph/0008114} {arXiv:hep-ph/0008114} \BibitemShut {NoStop}%
\bibitem [{\citenamefont {Bernabeu}\ \emph {et~al.}(2002)\citenamefont {Bernabeu}, \citenamefont {Papavassiliou},\ and\ \citenamefont {Vidal}}]{Bernabeu:2002nw}%
  \BibitemOpen
  \bibfield  {author} {\bibinfo {author} {\bibfnamefont {J.}~\bibnamefont {Bernabeu}}, \bibinfo {author} {\bibfnamefont {J.}~\bibnamefont {Papavassiliou}}, \ and\ \bibinfo {author} {\bibfnamefont {J.}~\bibnamefont {Vidal}},\ }\href {\doibase 10.1103/PhysRevLett.89.101802} {\bibfield  {journal} {\bibinfo  {journal} {Phys. Rev. Lett.}\ }\textbf {\bibinfo {volume} {89}},\ \bibinfo {pages} {101802} (\bibinfo {year} {2002})},\ \bibinfo {note} {[Erratum: Phys.Rev.Lett. 89, 229902 (2002)]},\ \Eprint {http://arxiv.org/abs/hep-ph/0206015} {arXiv:hep-ph/0206015} \BibitemShut {NoStop}%
\bibitem [{\citenamefont {Helm}(1956)}]{Helm:1956zz}%
  \BibitemOpen
  \bibfield  {author} {\bibinfo {author} {\bibfnamefont {R.~H.}\ \bibnamefont {Helm}},\ }\href {\doibase 10.1103/PhysRev.104.1466} {\bibfield  {journal} {\bibinfo  {journal} {Phys. Rev.}\ }\textbf {\bibinfo {volume} {104}},\ \bibinfo {pages} {1466} (\bibinfo {year} {1956})}\BibitemShut {NoStop}%
\bibitem [{\citenamefont {Piekarewicz}\ \emph {et~al.}(2016)\citenamefont {Piekarewicz}, \citenamefont {Linero}, \citenamefont {Giuliani},\ and\ \citenamefont {Chicken}}]{Piekarewicz:2016vbn}%
  \BibitemOpen
  \bibfield  {author} {\bibinfo {author} {\bibfnamefont {J.}~\bibnamefont {Piekarewicz}}, \bibinfo {author} {\bibfnamefont {A.~R.}\ \bibnamefont {Linero}}, \bibinfo {author} {\bibfnamefont {P.}~\bibnamefont {Giuliani}}, \ and\ \bibinfo {author} {\bibfnamefont {E.}~\bibnamefont {Chicken}},\ }\href {\doibase 10.1103/PhysRevC.94.034316} {\bibfield  {journal} {\bibinfo  {journal} {Phys. Rev. C}\ }\textbf {\bibinfo {volume} {94}},\ \bibinfo {pages} {034316} (\bibinfo {year} {2016})},\ \Eprint {http://arxiv.org/abs/1604.07799} {arXiv:1604.07799 [nucl-th]} \BibitemShut {NoStop}%
\bibitem [{\citenamefont {Klein}\ and\ \citenamefont {Nystrand}(1999)}]{Klein:1999qj}%
  \BibitemOpen
  \bibfield  {author} {\bibinfo {author} {\bibfnamefont {S.}~\bibnamefont {Klein}}\ and\ \bibinfo {author} {\bibfnamefont {J.}~\bibnamefont {Nystrand}},\ }\href {\doibase 10.1103/PhysRevC.60.014903} {\bibfield  {journal} {\bibinfo  {journal} {Phys. Rev. C}\ }\textbf {\bibinfo {volume} {60}},\ \bibinfo {pages} {014903} (\bibinfo {year} {1999})},\ \Eprint {http://arxiv.org/abs/hep-ph/9902259} {arXiv:hep-ph/9902259} \BibitemShut {NoStop}%
\bibitem [{\citenamefont {Fricke}\ \emph {et~al.}(1995)\citenamefont {Fricke}, \citenamefont {Bernhardt}, \citenamefont {Heilig}, \citenamefont {Schaller}, \citenamefont {Schellenberg}, \citenamefont {Shera},\ and\ \citenamefont {de~Jager}}]{Fricke:1995zz}%
  \BibitemOpen
  \bibfield  {author} {\bibinfo {author} {\bibfnamefont {G.}~\bibnamefont {Fricke}}, \bibinfo {author} {\bibfnamefont {C.}~\bibnamefont {Bernhardt}}, \bibinfo {author} {\bibfnamefont {K.}~\bibnamefont {Heilig}}, \bibinfo {author} {\bibfnamefont {L.~A.}\ \bibnamefont {Schaller}}, \bibinfo {author} {\bibfnamefont {L.}~\bibnamefont {Schellenberg}}, \bibinfo {author} {\bibfnamefont {E.~B.}\ \bibnamefont {Shera}}, \ and\ \bibinfo {author} {\bibfnamefont {C.~W.}\ \bibnamefont {de~Jager}},\ }\href {\doibase 10.1006/adnd.1995.1007} {\bibfield  {journal} {\bibinfo  {journal} {Atom. Data Nucl. Data Tabl.}\ }\textbf {\bibinfo {volume} {60}},\ \bibinfo {pages} {177} (\bibinfo {year} {1995})}\BibitemShut {NoStop}%
\bibitem [{\citenamefont {Fricke}\ and\ \citenamefont {Heilig}(2004)}]{Fricke2004}%
  \BibitemOpen
  \bibfield  {author} {\bibinfo {author} {\bibfnamefont {G.}~\bibnamefont {Fricke}}\ and\ \bibinfo {author} {\bibfnamefont {K.}~\bibnamefont {Heilig}},\ }\href {\doibase 10.1007/10856314_34} {\enquote {\bibinfo {title} {Nuclear charge radii $^{32}$ge germanium: Datasheet from landolt-bornstein - group i elementary particles, nuclei and atoms, volume 20},}\ } (\bibinfo {year} {2004}),\ \bibinfo {note} {copyright 2004 Springer-Verlag Berlin Heidelberg}\BibitemShut {NoStop}%
\bibitem [{\citenamefont {Angeli}\ and\ \citenamefont {Marinova}(2013)}]{Angeli:2013epw}%
  \BibitemOpen
  \bibfield  {author} {\bibinfo {author} {\bibfnamefont {I.}~\bibnamefont {Angeli}}\ and\ \bibinfo {author} {\bibfnamefont {K.~P.}\ \bibnamefont {Marinova}},\ }\href {\doibase 10.1016/j.adt.2011.12.006} {\bibfield  {journal} {\bibinfo  {journal} {Atom. Data Nucl. Data Tabl.}\ }\textbf {\bibinfo {volume} {99}},\ \bibinfo {pages} {69} (\bibinfo {year} {2013})}\BibitemShut {NoStop}%
\bibitem [{\citenamefont {Berger}\ \emph {et~al.}(1991)\citenamefont {Berger}, \citenamefont {Girod},\ and\ \citenamefont {Gogny}}]{Berger:1991zza}%
  \BibitemOpen
  \bibfield  {author} {\bibinfo {author} {\bibfnamefont {J.~F.}\ \bibnamefont {Berger}}, \bibinfo {author} {\bibfnamefont {M.}~\bibnamefont {Girod}}, \ and\ \bibinfo {author} {\bibfnamefont {D.}~\bibnamefont {Gogny}},\ }\href {\doibase 10.1016/0010-4655(91)90263-K} {\bibfield  {journal} {\bibinfo  {journal} {Comput. Phys. Commun.}\ }\textbf {\bibinfo {volume} {63}},\ \bibinfo {pages} {365} (\bibinfo {year} {1991})}\BibitemShut {NoStop}%
\bibitem [{\citenamefont {Chabanat}\ \emph {et~al.}(1998)\citenamefont {Chabanat}, \citenamefont {Bonche}, \citenamefont {Haensel}, \citenamefont {Meyer},\ and\ \citenamefont {Schaeffer}}]{Chabanat:1997un}%
  \BibitemOpen
  \bibfield  {author} {\bibinfo {author} {\bibfnamefont {E.}~\bibnamefont {Chabanat}}, \bibinfo {author} {\bibfnamefont {P.}~\bibnamefont {Bonche}}, \bibinfo {author} {\bibfnamefont {P.}~\bibnamefont {Haensel}}, \bibinfo {author} {\bibfnamefont {J.}~\bibnamefont {Meyer}}, \ and\ \bibinfo {author} {\bibfnamefont {R.}~\bibnamefont {Schaeffer}},\ }\href {\doibase 10.1016/S0375-9474(98)00180-8} {\bibfield  {journal} {\bibinfo  {journal} {Nucl. Phys. A}\ }\textbf {\bibinfo {volume} {635}},\ \bibinfo {pages} {231} (\bibinfo {year} {1998})},\ \bibinfo {note} {[Erratum: Nucl.Phys.A 643, 441--441 (1998)]}\BibitemShut {NoStop}%
\bibitem [{\citenamefont {Hoferichter}\ \emph {et~al.}(2020)\citenamefont {Hoferichter}, \citenamefont {Men\'endez},\ and\ \citenamefont {Schwenk}}]{Hoferichter:2020osn}%
  \BibitemOpen
  \bibfield  {author} {\bibinfo {author} {\bibfnamefont {M.}~\bibnamefont {Hoferichter}}, \bibinfo {author} {\bibfnamefont {J.}~\bibnamefont {Men\'endez}}, \ and\ \bibinfo {author} {\bibfnamefont {A.}~\bibnamefont {Schwenk}},\ }\href {\doibase 10.1103/PhysRevD.102.074018} {\bibfield  {journal} {\bibinfo  {journal} {Phys. Rev. D}\ }\textbf {\bibinfo {volume} {102}},\ \bibinfo {pages} {074018} (\bibinfo {year} {2020})},\ \Eprint {http://arxiv.org/abs/2007.08529} {arXiv:2007.08529 [hep-ph]} \BibitemShut {NoStop}%
\bibitem [{\citenamefont {Khaleq}\ \emph {et~al.}(2025)\citenamefont {Khaleq}, \citenamefont {Newstead}, \citenamefont {Simenel},\ and\ \citenamefont {Stuchbery}}]{PhysRevD.111.033003}%
  \BibitemOpen
  \bibfield  {author} {\bibinfo {author} {\bibfnamefont {R.~A.}\ \bibnamefont {Khaleq}}, \bibinfo {author} {\bibfnamefont {J.~L.}\ \bibnamefont {Newstead}}, \bibinfo {author} {\bibfnamefont {C.}~\bibnamefont {Simenel}}, \ and\ \bibinfo {author} {\bibfnamefont {A.~E.}\ \bibnamefont {Stuchbery}},\ }\href {\doibase 10.1103/PhysRevD.111.033003} {\bibfield  {journal} {\bibinfo  {journal} {Phys. Rev. D}\ }\textbf {\bibinfo {volume} {111}},\ \bibinfo {pages} {033003} (\bibinfo {year} {2025})}\BibitemShut {NoStop}%
\bibitem [{\citenamefont {Co'}\ \emph {et~al.}(2020)\citenamefont {Co'}, \citenamefont {Anguiano},\ and\ \citenamefont {Lallena}}]{Co:2020gwl}%
  \BibitemOpen
  \bibfield  {author} {\bibinfo {author} {\bibfnamefont {G.}~\bibnamefont {Co'}}, \bibinfo {author} {\bibfnamefont {M.}~\bibnamefont {Anguiano}}, \ and\ \bibinfo {author} {\bibfnamefont {A.~M.}\ \bibnamefont {Lallena}},\ }\href {\doibase 10.1088/1475-7516/2020/04/044} {\bibfield  {journal} {\bibinfo  {journal} {JCAP}\ }\textbf {\bibinfo {volume} {04}},\ \bibinfo {pages} {044} (\bibinfo {year} {2020})},\ \Eprint {http://arxiv.org/abs/2001.04684} {arXiv:2001.04684 [nucl-th]} \BibitemShut {NoStop}%
\bibitem [{\citenamefont {Co}\ \emph {et~al.}(2021)\citenamefont {Co}, \citenamefont {Anguiano},\ and\ \citenamefont {Lallena}}]{Co:2021ijy}%
  \BibitemOpen
  \bibfield  {author} {\bibinfo {author} {\bibfnamefont {G.}~\bibnamefont {Co}}, \bibinfo {author} {\bibfnamefont {M.}~\bibnamefont {Anguiano}}, \ and\ \bibinfo {author} {\bibfnamefont {A.~M.}\ \bibnamefont {Lallena}},\ }\href {\doibase 10.1103/PhysRevC.104.014313} {\bibfield  {journal} {\bibinfo  {journal} {Phys. Rev. C}\ }\textbf {\bibinfo {volume} {104}},\ \bibinfo {pages} {014313} (\bibinfo {year} {2021})},\ \Eprint {http://arxiv.org/abs/2107.09938} {arXiv:2107.09938 [nucl-th]} \BibitemShut {NoStop}%
\bibitem [{\citenamefont {Atzori~Corona}\ \emph {et~al.}(2025{\natexlab{b}})\citenamefont {Atzori~Corona}, \citenamefont {Cadeddu}, \citenamefont {Cargioli}, \citenamefont {Dordei},\ and\ \citenamefont {Giunti}}]{AtzoriCorona:2025ygn}%
  \BibitemOpen
  \bibfield  {author} {\bibinfo {author} {\bibfnamefont {M.}~\bibnamefont {Atzori~Corona}}, \bibinfo {author} {\bibfnamefont {M.}~\bibnamefont {Cadeddu}}, \bibinfo {author} {\bibfnamefont {N.}~\bibnamefont {Cargioli}}, \bibinfo {author} {\bibfnamefont {F.}~\bibnamefont {Dordei}}, \ and\ \bibinfo {author} {\bibfnamefont {C.}~\bibnamefont {Giunti}},\ }\href@noop {} {\  (\bibinfo {year} {2025}{\natexlab{b}})},\ \Eprint {http://arxiv.org/abs/2501.18550} {arXiv:2501.18550 [hep-ph]} \BibitemShut {NoStop}%
\bibitem [{\citenamefont {Periss\'e}\ \emph {et~al.}(2023)\citenamefont {Periss\'e}, \citenamefont {Onillon}, \citenamefont {Mougeot}, \citenamefont {Vivier}, \citenamefont {Lasserre}, \citenamefont {Letourneau}, \citenamefont {Lhuillier},\ and\ \citenamefont {Mention}}]{Perisse:2023efm}%
  \BibitemOpen
  \bibfield  {author} {\bibinfo {author} {\bibfnamefont {L.}~\bibnamefont {Periss\'e}}, \bibinfo {author} {\bibfnamefont {A.}~\bibnamefont {Onillon}}, \bibinfo {author} {\bibfnamefont {X.}~\bibnamefont {Mougeot}}, \bibinfo {author} {\bibfnamefont {M.}~\bibnamefont {Vivier}}, \bibinfo {author} {\bibfnamefont {T.}~\bibnamefont {Lasserre}}, \bibinfo {author} {\bibfnamefont {A.}~\bibnamefont {Letourneau}}, \bibinfo {author} {\bibfnamefont {D.}~\bibnamefont {Lhuillier}}, \ and\ \bibinfo {author} {\bibfnamefont {G.}~\bibnamefont {Mention}},\ }\href {\doibase 10.1103/PhysRevC.108.055501} {\bibfield  {journal} {\bibinfo  {journal} {Phys. Rev. C}\ }\textbf {\bibinfo {volume} {108}},\ \bibinfo {pages} {055501} (\bibinfo {year} {2023})},\ \Eprint {http://arxiv.org/abs/2304.14992} {arXiv:2304.14992 [nucl-ex]} \BibitemShut {NoStop}%
\bibitem [{\citenamefont {Akimov}\ \emph {et~al.}(2021{\natexlab{b}})\citenamefont {Akimov} \emph {et~al.}}]{Akimov:2021dab}%
  \BibitemOpen
  \bibfield  {author} {\bibinfo {author} {\bibfnamefont {D.}~\bibnamefont {Akimov}} \emph {et~al.},\ }\href@noop {} {\enquote {\bibinfo {title} {{Measurement of the Coherent Elastic Neutrino-Nucleus Scattering Cross Section on CsI by COHERENT}},}\ } (\bibinfo {year} {2021}{\natexlab{b}}),\ \Eprint {http://arxiv.org/abs/2110.07730} {arXiv:2110.07730 [hep-ex]} \BibitemShut {NoStop}%
\bibitem [{\citenamefont {Akimov}\ \emph {et~al.}(2020)\citenamefont {Akimov} \emph {et~al.}}]{COHERENT:2020ybo}%
  \BibitemOpen
  \bibfield  {author} {\bibinfo {author} {\bibfnamefont {D.}~\bibnamefont {Akimov}} \emph {et~al.} (\bibinfo {collaboration} {COHERENT}),\ }\href {\doibase 10.5281/zenodo.3903810} {\enquote {\bibinfo {title} {{COHERENT Collaboration data release from the first detection of coherent elastic neutrino-nucleus scattering on argon}},}\ } (\bibinfo {year} {2020}),\ \Eprint {http://arxiv.org/abs/2006.12659} {arXiv:2006.12659 [nucl-ex]} \BibitemShut {NoStop}%
\bibitem [{\citenamefont {Colaresi}\ \emph {et~al.}(2022)\citenamefont {Colaresi}, \citenamefont {Collar}, \citenamefont {Hossbach}, \citenamefont {Lewis},\ and\ \citenamefont {Yocum}}]{Colaresi:2022obx}%
  \BibitemOpen
  \bibfield  {author} {\bibinfo {author} {\bibfnamefont {J.}~\bibnamefont {Colaresi}}, \bibinfo {author} {\bibfnamefont {J.~I.}\ \bibnamefont {Collar}}, \bibinfo {author} {\bibfnamefont {T.~W.}\ \bibnamefont {Hossbach}}, \bibinfo {author} {\bibfnamefont {C.~M.}\ \bibnamefont {Lewis}}, \ and\ \bibinfo {author} {\bibfnamefont {K.~M.}\ \bibnamefont {Yocum}},\ }\href {\doibase 10.1103/PhysRevLett.129.211802} {\bibfield  {journal} {\bibinfo  {journal} {Phys. Rev. Lett.}\ }\textbf {\bibinfo {volume} {129}},\ \bibinfo {pages} {211802} (\bibinfo {year} {2022})},\ \Eprint {http://arxiv.org/abs/2202.09672} {arXiv:2202.09672 [hep-ex]} \BibitemShut {NoStop}%
\bibitem [{\citenamefont {Atzori~Corona}\ \emph {et~al.}(2024{\natexlab{b}})\citenamefont {Atzori~Corona}, \citenamefont {Cadeddu}, \citenamefont {Cargioli}, \citenamefont {Dordei},\ and\ \citenamefont {Giunti}}]{AtzoriCorona:2023ais}%
  \BibitemOpen
  \bibfield  {author} {\bibinfo {author} {\bibfnamefont {M.}~\bibnamefont {Atzori~Corona}}, \bibinfo {author} {\bibfnamefont {M.}~\bibnamefont {Cadeddu}}, \bibinfo {author} {\bibfnamefont {N.}~\bibnamefont {Cargioli}}, \bibinfo {author} {\bibfnamefont {F.}~\bibnamefont {Dordei}}, \ and\ \bibinfo {author} {\bibfnamefont {C.}~\bibnamefont {Giunti}},\ }\href {\doibase 10.1016/j.physletb.2024.138627} {\bibfield  {journal} {\bibinfo  {journal} {Phys. Lett. B}\ }\textbf {\bibinfo {volume} {852}},\ \bibinfo {pages} {138627} (\bibinfo {year} {2024}{\natexlab{b}})},\ \Eprint {http://arxiv.org/abs/2307.12911} {arXiv:2307.12911 [hep-ph]} \BibitemShut {NoStop}%
\bibitem [{\citenamefont {Li}\ \emph {et~al.}(2025)\citenamefont {Li}, \citenamefont {Herrera},\ and\ \citenamefont {Huber}}]{Li:2025pfw}%
  \BibitemOpen
  \bibfield  {author} {\bibinfo {author} {\bibfnamefont {Y.}~\bibnamefont {Li}}, \bibinfo {author} {\bibfnamefont {G.}~\bibnamefont {Herrera}}, \ and\ \bibinfo {author} {\bibfnamefont {P.}~\bibnamefont {Huber}},\ }\href@noop {} {\  (\bibinfo {year} {2025})},\ \Eprint {http://arxiv.org/abs/2502.12308} {arXiv:2502.12308 [hep-ph]} \BibitemShut {NoStop}%
\bibitem [{\citenamefont {Lindhard}\ \emph {et~al.}(1963)\citenamefont {Lindhard}, \citenamefont {Nielsen}, \citenamefont {Scharff},\ and\ \citenamefont {Thomsen}}]{Lindhard_theo}%
  \BibitemOpen
  \bibfield  {author} {\bibinfo {author} {\bibfnamefont {J.}~\bibnamefont {Lindhard}}, \bibinfo {author} {\bibfnamefont {V.}~\bibnamefont {Nielsen}}, \bibinfo {author} {\bibfnamefont {M.}~\bibnamefont {Scharff}}, \ and\ \bibinfo {author} {\bibfnamefont {P.~V.}\ \bibnamefont {Thomsen}},\ }\href {https://www.osti.gov/biblio/4701226} {\bibfield  {journal} {\bibinfo  {journal} {Kgl. Danske Videnskab., Selskab. Mat. Fys. Medd.}\ }\textbf {\bibinfo {volume} {33, 10}} (\bibinfo {year} {1963})}\BibitemShut {NoStop}%
\bibitem [{\citenamefont {Bonhomme}\ \emph {et~al.}(2022)\citenamefont {Bonhomme} \emph {et~al.}}]{Bonhomme:2022lcz}%
  \BibitemOpen
  \bibfield  {author} {\bibinfo {author} {\bibfnamefont {A.}~\bibnamefont {Bonhomme}} \emph {et~al.},\ }\href {\doibase 10.1140/epjc/s10052-022-10768-1} {\bibfield  {journal} {\bibinfo  {journal} {Eur. Phys. J. C}\ }\textbf {\bibinfo {volume} {82}},\ \bibinfo {pages} {815} (\bibinfo {year} {2022})},\ \Eprint {http://arxiv.org/abs/2202.03754} {arXiv:2202.03754 [physics.ins-det]} \BibitemShut {NoStop}%
\bibitem [{\citenamefont {Bouabid}(2024)}]{M7Germanio}%
  \BibitemOpen
  \bibfield  {author} {\bibinfo {author} {\bibfnamefont {R.}~\bibnamefont {Bouabid}} (\bibinfo {collaboration} {COHERENT}),\ }\href {https://indico.global/event/6083/contributions/50001/attachments/24943/42694/m7s_rbouabid_upload.pdf} {\enquote {\bibinfo {title} {{First Measurement of CEvNS on Germanium by COHERENT}},}\ } (\bibinfo {year} {2024}),\ \bibinfo {note} {presentation at the Magnificent CENNS Workshop}\BibitemShut {NoStop}%
\bibitem [{\citenamefont {Atzori~Corona}\ \emph {et~al.}(2024{\natexlab{c}})\citenamefont {Atzori~Corona}, \citenamefont {Cadeddu}, \citenamefont {Cargioli}, \citenamefont {Dordei},\ and\ \citenamefont {Giunti}}]{AtzoriCorona:2024vhj}%
  \BibitemOpen
  \bibfield  {author} {\bibinfo {author} {\bibfnamefont {M.}~\bibnamefont {Atzori~Corona}}, \bibinfo {author} {\bibfnamefont {M.}~\bibnamefont {Cadeddu}}, \bibinfo {author} {\bibfnamefont {N.}~\bibnamefont {Cargioli}}, \bibinfo {author} {\bibfnamefont {F.}~\bibnamefont {Dordei}}, \ and\ \bibinfo {author} {\bibfnamefont {C.}~\bibnamefont {Giunti}},\ }\href {\doibase 10.1103/PhysRevD.110.033005} {\bibfield  {journal} {\bibinfo  {journal} {Phys. Rev. D}\ }\textbf {\bibinfo {volume} {110}},\ \bibinfo {pages} {033005} (\bibinfo {year} {2024}{\natexlab{c}})},\ \Eprint {http://arxiv.org/abs/2405.09416} {arXiv:2405.09416 [hep-ph]} \BibitemShut {NoStop}%
\bibitem [{\citenamefont {Cadeddu}\ \emph {et~al.}(2024)\citenamefont {Cadeddu}, \citenamefont {Cargioli}, \citenamefont {Erler}, \citenamefont {Gorchtein}, \citenamefont {Piekarewicz}, \citenamefont {Roca-Maza},\ and\ \citenamefont {Spiesberger}}]{Cadeddu:2024baq}%
  \BibitemOpen
  \bibfield  {author} {\bibinfo {author} {\bibfnamefont {M.}~\bibnamefont {Cadeddu}}, \bibinfo {author} {\bibfnamefont {N.}~\bibnamefont {Cargioli}}, \bibinfo {author} {\bibfnamefont {J.}~\bibnamefont {Erler}}, \bibinfo {author} {\bibfnamefont {M.}~\bibnamefont {Gorchtein}}, \bibinfo {author} {\bibfnamefont {J.}~\bibnamefont {Piekarewicz}}, \bibinfo {author} {\bibfnamefont {X.}~\bibnamefont {Roca-Maza}}, \ and\ \bibinfo {author} {\bibfnamefont {H.}~\bibnamefont {Spiesberger}},\ }\href {\doibase 10.1103/PhysRevC.110.035501} {\bibfield  {journal} {\bibinfo  {journal} {Phys. Rev. C}\ }\textbf {\bibinfo {volume} {110}},\ \bibinfo {pages} {035501} (\bibinfo {year} {2024})},\ \Eprint {http://arxiv.org/abs/2407.09743} {arXiv:2407.09743 [hep-ph]} \BibitemShut {NoStop}%
\bibitem [{\citenamefont {Deniz}\ \emph {et~al.}(2010)\citenamefont {Deniz} \emph {et~al.}}]{TEXONO:2009knm}%
  \BibitemOpen
  \bibfield  {author} {\bibinfo {author} {\bibfnamefont {M.}~\bibnamefont {Deniz}} \emph {et~al.} (\bibinfo {collaboration} {TEXONO}),\ }\href {\doibase 10.1103/PhysRevD.81.072001} {\bibfield  {journal} {\bibinfo  {journal} {Phys. Rev. D}\ }\textbf {\bibinfo {volume} {81}},\ \bibinfo {pages} {072001} (\bibinfo {year} {2010})},\ \Eprint {http://arxiv.org/abs/0911.1597} {arXiv:0911.1597 [hep-ex]} \BibitemShut {NoStop}%
\bibitem [{\citenamefont {Auerbach}\ \emph {et~al.}(2001)\citenamefont {Auerbach} \emph {et~al.}}]{LSND:2001akn}%
  \BibitemOpen
  \bibfield  {author} {\bibinfo {author} {\bibfnamefont {L.~B.}\ \bibnamefont {Auerbach}} \emph {et~al.} (\bibinfo {collaboration} {LSND}),\ }\href {\doibase 10.1103/PhysRevD.63.112001} {\bibfield  {journal} {\bibinfo  {journal} {Phys. Rev. D}\ }\textbf {\bibinfo {volume} {63}},\ \bibinfo {pages} {112001} (\bibinfo {year} {2001})},\ \Eprint {http://arxiv.org/abs/hep-ex/0101039} {arXiv:hep-ex/0101039} \BibitemShut {NoStop}%
\bibitem [{\citenamefont {Allen}\ \emph {et~al.}(1993)\citenamefont {Allen}, \citenamefont {Chen}, \citenamefont {Doe}, \citenamefont {Hausammann}, \citenamefont {Lee}, \citenamefont {Lu}, \citenamefont {Mahler}, \citenamefont {Potter}, \citenamefont {Wang}, \citenamefont {Bowles}, \citenamefont {Burman}, \citenamefont {Carlini}, \citenamefont {Cochran}, \citenamefont {Frank}, \citenamefont {Piasetzky}, \citenamefont {Sandberg}, \citenamefont {Krakauer},\ and\ \citenamefont {Talaga}}]{Allen:1992qe}%
  \BibitemOpen
  \bibfield  {author} {\bibinfo {author} {\bibfnamefont {R.~C.}\ \bibnamefont {Allen}}, \bibinfo {author} {\bibfnamefont {H.~H.}\ \bibnamefont {Chen}}, \bibinfo {author} {\bibfnamefont {P.~J.}\ \bibnamefont {Doe}}, \bibinfo {author} {\bibfnamefont {R.}~\bibnamefont {Hausammann}}, \bibinfo {author} {\bibfnamefont {W.~P.}\ \bibnamefont {Lee}}, \bibinfo {author} {\bibfnamefont {X.~Q.}\ \bibnamefont {Lu}}, \bibinfo {author} {\bibfnamefont {H.~J.}\ \bibnamefont {Mahler}}, \bibinfo {author} {\bibfnamefont {M.~E.}\ \bibnamefont {Potter}}, \bibinfo {author} {\bibfnamefont {K.~C.}\ \bibnamefont {Wang}}, \bibinfo {author} {\bibfnamefont {T.~J.}\ \bibnamefont {Bowles}}, \bibinfo {author} {\bibfnamefont {R.~L.}\ \bibnamefont {Burman}}, \bibinfo {author} {\bibfnamefont {R.~D.}\ \bibnamefont {Carlini}}, \bibinfo {author} {\bibfnamefont {D.~R.~F.}\ \bibnamefont {Cochran}}, \bibinfo {author} {\bibfnamefont {J.~S.}\ \bibnamefont {Frank}}, \bibinfo {author} {\bibfnamefont {E.}~\bibnamefont {Piasetzky}}, \bibinfo {author}
  {\bibfnamefont {V.~D.}\ \bibnamefont {Sandberg}}, \bibinfo {author} {\bibfnamefont {D.~A.}\ \bibnamefont {Krakauer}}, \ and\ \bibinfo {author} {\bibfnamefont {R.~L.}\ \bibnamefont {Talaga}},\ }\href {\doibase 10.1103/PhysRevD.47.11} {\bibfield  {journal} {\bibinfo  {journal} {Phys. Rev. D}\ }\textbf {\bibinfo {volume} {47}},\ \bibinfo {pages} {11} (\bibinfo {year} {1993})}\BibitemShut {NoStop}%
\bibitem [{\citenamefont {Aalbers}\ \emph {et~al.}(2023)\citenamefont {Aalbers} \emph {et~al.}}]{LZ:2022lsv}%
  \BibitemOpen
  \bibfield  {author} {\bibinfo {author} {\bibfnamefont {J.}~\bibnamefont {Aalbers}} \emph {et~al.} (\bibinfo {collaboration} {LZ}),\ }\href {\doibase 10.1103/PhysRevLett.131.041002} {\bibfield  {journal} {\bibinfo  {journal} {Phys. Rev. Lett.}\ }\textbf {\bibinfo {volume} {131}},\ \bibinfo {pages} {041002} (\bibinfo {year} {2023})},\ \Eprint {http://arxiv.org/abs/2207.03764} {arXiv:2207.03764 [hep-ex]} \BibitemShut {NoStop}%
\bibitem [{\citenamefont {Zhang}\ \emph {et~al.}(2022)\citenamefont {Zhang} \emph {et~al.}}]{PandaX:2022ood}%
  \BibitemOpen
  \bibfield  {author} {\bibinfo {author} {\bibfnamefont {D.}~\bibnamefont {Zhang}} \emph {et~al.} (\bibinfo {collaboration} {PandaX}),\ }\href {\doibase 10.1103/PhysRevLett.129.161804} {\bibfield  {journal} {\bibinfo  {journal} {Phys. Rev. Lett.}\ }\textbf {\bibinfo {volume} {129}},\ \bibinfo {pages} {161804} (\bibinfo {year} {2022})},\ \Eprint {http://arxiv.org/abs/2206.02339} {arXiv:2206.02339 [hep-ex]} \BibitemShut {NoStop}%
\bibitem [{\citenamefont {Aprile}\ \emph {et~al.}(2022)\citenamefont {Aprile} \emph {et~al.}}]{XENON:2022ltv}%
  \BibitemOpen
  \bibfield  {author} {\bibinfo {author} {\bibfnamefont {E.}~\bibnamefont {Aprile}} \emph {et~al.} (\bibinfo {collaboration} {XENON}),\ }\href {\doibase 10.1103/PhysRevLett.129.161805} {\bibfield  {journal} {\bibinfo  {journal} {Phys. Rev. Lett.}\ }\textbf {\bibinfo {volume} {129}},\ \bibinfo {pages} {161805} (\bibinfo {year} {2022})},\ \Eprint {http://arxiv.org/abs/2207.11330} {arXiv:2207.11330 [hep-ex]} \BibitemShut {NoStop}%
\bibitem [{\citenamefont {Akimov}\ \emph {et~al.}(2021{\natexlab{c}})\citenamefont {Akimov} \emph {et~al.}}]{COHERENT:2021xhx}%
  \BibitemOpen
  \bibfield  {author} {\bibinfo {author} {\bibfnamefont {D.}~\bibnamefont {Akimov}} \emph {et~al.} (\bibinfo {collaboration} {COHERENT}),\ }\href {\doibase 10.1088/1748-0221/16/08/P08048} {\bibfield  {journal} {\bibinfo  {journal} {JINST}\ }\textbf {\bibinfo {volume} {16}},\ \bibinfo {pages} {P08048} (\bibinfo {year} {2021}{\natexlab{c}})},\ \Eprint {http://arxiv.org/abs/2104.09605} {arXiv:2104.09605 [physics.ins-det]} \BibitemShut {NoStop}%
\end{thebibliography}%

\end{document}